\documentclass[12pt]{article}
\usepackage[letterpaper,margin=1in]{geometry}
\usepackage{newtxtext,newtxmath}
\usepackage{microtype}
\usepackage[normalem]{ulem}
\usepackage{amsmath}
\usepackage{graphicx}
\usepackage{xcolor}
\usepackage[normalem]{ulem}
\usepackage{soul}
\usepackage[hidelinks]{hyperref}
\usepackage{url}
\usepackage[numbers,sort&compress]{natbib}
\usepackage{setspace}
\usepackage[font=small,labelfont=bf]{caption}

\newcommand{\scititle}{Femtosecond laser-induced cavitation seeds liquid-jet breakup beyond the Rayleigh–Plateau stability limit}

\title{\Large\bfseries\scititle}

\author{
Sourabh Singh$^{1}$,
Tamanna$^{1}$,
S.~Sree~Harsha$^{1}$,
Prashant Kumar Singh$^{1,*}$\\[1em]
$^{1}$Tata Institute of Fundamental Research Hyderabad,\\
36/P Gopanpally, Hyderabad 500046, Telangana, India\\[0.5em]
$^{*}$Correspondence: \href{mailto:pks@tifrh.res.in}{pks@tifrh.res.in}
}

\date{}

\begin{document}


\maketitle

\begin{abstract}
The Rayleigh–Plateau instability predicts that infinitesimal perturbations of a liquid jet grow only for $kR<1$, with $kR=1$ marking the classical stability boundary. Here, we show that femtosecond-laser-induced cavitation generates a localized, impulsive recoil perturbation on a flowing liquid jet, enabling a two-dimensional map of jet breakup across the $(kR,\epsilon/R)$ plane. The perturbation's spacing (wavelength) and strength (amplitude) are set independently by the laser repetition rate and pulse energy, respectively. At low amplitudes, tuning the repetition rate recovers the classical Rayleigh--Plateau dispersion relation and its \(kR=1\) cutoff, while increasing the wavelength enables controlled generation of monodisperse droplet trains, bidisperse droplet populations, and ultimately isolated single droplets. Along the amplitude axis, low pulse energies produce a seed that amplifies only the imposed wavenumber, whereas above a critical energy $E_c$, the seed broadens to excite many wavenumbers simultaneously. At still higher amplitudes, we demonstrate deterministic breakup for $kR>1$, extending to $kR=1.9$, with laser-synchronized droplet production rates up to 0.2~MHz. These results establish localized femtosecond-laser-induced cavitation as a route to controlled Rayleigh--Plateau breakup both within and beyond the classical linear-stability boundary.
\end{abstract}


\section*{Introduction}

Elongated streams of matter, held together by cohesive forces, appear across an extraordinary range of scales, from thermally fluctuating nanometer-scale jets~\cite{NanoJet_rayleigh_Science2000, Nanojet_PRL_2002} to biological filaments~\cite{Setru_NatPhys2021, platelet_Rayleigh_PNAS_2020} and gravitational analogues~\cite{Cardoso_PRL2006, Black_string_PRL_2010, Chen_PRL2025}. Yet, the very forces that stabilize these structures ultimately drive breakup by reducing the total interfacial energy. This counterintuitive phenomenon, known as the Rayleigh–Plateau instability, originally developed to describe the breakup of liquid columns into droplets~\cite{Plateau1857, Rayleigh1878, Rayleigh1892_minimal}, is one of the fundamental paradigms of hydrodynamic instability and singularity formation \cite{singularity_Axisymetric_PRL_1992, singularity_Axisymetric_PRL_1993, Cascade_Science_1994}.  Examples range from the breakup of liquid jets into droplets~\cite{Plateau1857, Rayleigh1878, Rayleigh1892_minimal, Eggers_RMP1997, Eggers_RPP2008} and granular streams into clusters~\cite{Jaeger_Nature2009, Prado_PRL2011}, from the breakup of liquid-metal streams~\cite{Song_PNAS2020}, to the fragmentation of biological filaments~\cite{Setru_NatPhys2021, platelet_Rayleigh_PNAS_2020},  and even gravitational analogues of black strings~\cite{Cardoso_PRL2006, Black_string_PRL_2010, Chen_PRL2025}. 

Classical linear stability theory predicts that perturbations of a cylindrical liquid jet with dimensionless wavenumber \(kR<1\) amplify exponentially, whereas shorter-wavelength modes decay~\cite{Chandrasekhar1961,Eggers_RMP1997}. For more than a century, this framework has formed the basis for technologies ranging from atomization~\cite{Orme1991, Heister1997} and inkjet printing~\cite{Basaran2002, Derby2010} to microfluidics~\cite{Zhu2017, Garstecki2006,Anna2003}. Controlling this breakup requires imposing well-defined perturbations, prompting a variety of mechanical~\cite{Lindblad1965, Sweet1965, Schneider1964, Rodriguez2011}, acoustic~\cite{Donelly1966, Frommhold2014}, electrical~\cite{Hrdina1989EHD, BardeauIsolatedDrops}, and optical approaches~\cite{Zhao2021, Nahas1990, CHEN2025105182, Liu2021}.

Optical approaches provide contactless forcing and have been employed to control Rayleigh--Plateau breakup through waveguided axial coupling~\cite{ Liu2021, CHEN2025105182}, thermocapillary forcing~\cite{Zhao2021, Nahas1990}, and radiation-pressure feedback~\cite{Liu2021}. These studies established precise optical control over the perturbation wavelength and breakup frequency. In these approaches, however, the optical input is converted into a distributed perturbation through axial optical-pressure, temperature, or surface-tension fields. By contrast, laser-induced cavitation~\cite{Park2011} produces an impulsive and highly localized mechanical displacement of the liquid interface. Such impulsive forcing can access high-amplitude, nonlinear deformation regimes~\cite{Yuen1968,Nayfeh1970,Eggers_RPP2008}, in which finite-amplitude effects are known to shift the effective stability boundary away from $kR=1$~\cite{Ashgriz1995}. However, whether such shifts can drive complete droplet pinch-off at wavenumbers where infinitesimal perturbations are linearly stable, rather than merely produce transient, non-fragmenting deformation, has not been directly demonstrated experimentally.

Here, we use femtosecond laser pulses to directly write deterministic, tunable-amplitude surface deformations onto an initially cylindrical jet, via localized optical breakdown and cavitation-driven recoil~\cite{Cheng2025,Patrascioiu2014,Nakagawa2017,Li2018,Supponen2015,Kohsakowski2017,Han2022,Vogel2003,Sun2009,Tagawa2012,Chichkov1996, Gattass2008,Vogel2005, Rossello2022, Krivokorytov2018, Gonzalez2016, Robert2007, Benjamin1966}. The laser repetition rate and pulse energy provide tunable control of the perturbation spacing ($kR$) and initial amplitude ($\epsilon/R$), enabling systematic exploration of Rayleigh--Plateau dynamics in the two-dimensional parameter space $(kR,\epsilon/R)$. At low amplitudes, the measured growth rates recover the classical Rayleigh--Plateau dispersion relation and its $kR=1$ stability boundary. Increasing the perturbation amplitude drives a transition from selective narrow mode growth to broadband nonlinear evolution, with a sharp crossover at a critical energy $E_c$ beyond which the breakup length saturates. Most significantly, at high amplitudes we observe deterministic pinch-off even for $kR>1$, where infinitesimal perturbations are otherwise linearly stable. Varying the perturbation spacing reveals a progression from monodisperse droplet trains to bidisperse breakup and ultimately to isolated single droplets in the extreme long-wavelength limit ($kR\approx0.05$, $\lambda/D\approx60$), where widely spaced perturbations evolve independently as localized capillary waves. These regimes are mapped in the $(kR,\epsilon/R)$ parameter space, establishing localized femtosecond seeding as a route to control capillary breakup across linear, nonlinear, and beyond-stability regimes.

\color{black}

\section*{Result}
\subsection*{Cavitation-driven seeding of Rayleigh--Plateau perturbations}

We seed localized perturbations on a flowing liquid jet using a periodic, multi-kHz femtosecond laser. Optical breakdown generates plasma and shock dynamics on picosecond-to-nanosecond timescales; as shown below, these do not themselves seed a lasting deformation. Instead, cavitation-driven surface deformation on microsecond timescales provides the persistent perturbation from which Rayleigh--Plateau breakup develops~\cite{Femto_to_micro_Mazur, Eggers_RMP1997, Eggers_RPP2008, Mechanical_Vogel}. Time-resolved pump--probe shadowgraphy (Fig.~1A) resolves the complete evolution from plasma formation to droplet fragmentation (see Supplementary Material for breakdown and cavitation dynamics).

To generate these localized perturbations, the incident pulse is tightly focused inside the jet, ensuring spatially localized energy deposition near the rear liquid--air interface via refractive focusing from the cylindrical geometry~\cite{Caustic_Book, Cylinder_focusOL, SinghCaustic2026} (see Methods and Supplementary). At the pulse energies used, peak intensities exceed the threshold for optical breakdown in water, generating a micron-scale plasma that expands rapidly and launches a shock wave with an estimated peak pressure of several hundred MPa to ~1~GPa (Fig.~1B1--B2)~\cite{Optical_breakdown_Vogel, Optical_breakdown_Water, Gpa_Shock_Nat2016}. Despite the high pressures involved, the shock stage produces no measurable long-lived surface deformation (Fig.~1B3). The shock and subsequent acoustic waves evolve on nanosecond time scales, more than four orders of magnitude faster than the capillary--inertial response time ($\tau_{\rm shock}\sim3$~ns vs.\ $\tau_{\rm cap}\sim42~\mu$s). As a result, the shock-driven acoustic field undergoes multiple compression--reflection cycles within the jet without producing a persistent capillary deformation capable of seeding Rayleigh--Plateau growth~\cite{jet_shock_Nano2020}. The momentum delivered during the shock stage is insufficient to displace the interface by a measurable amount on its own (Supplementary Section~1.4).

\begin{figure}[h!]
\centering
\includegraphics[width=1.0\textwidth]{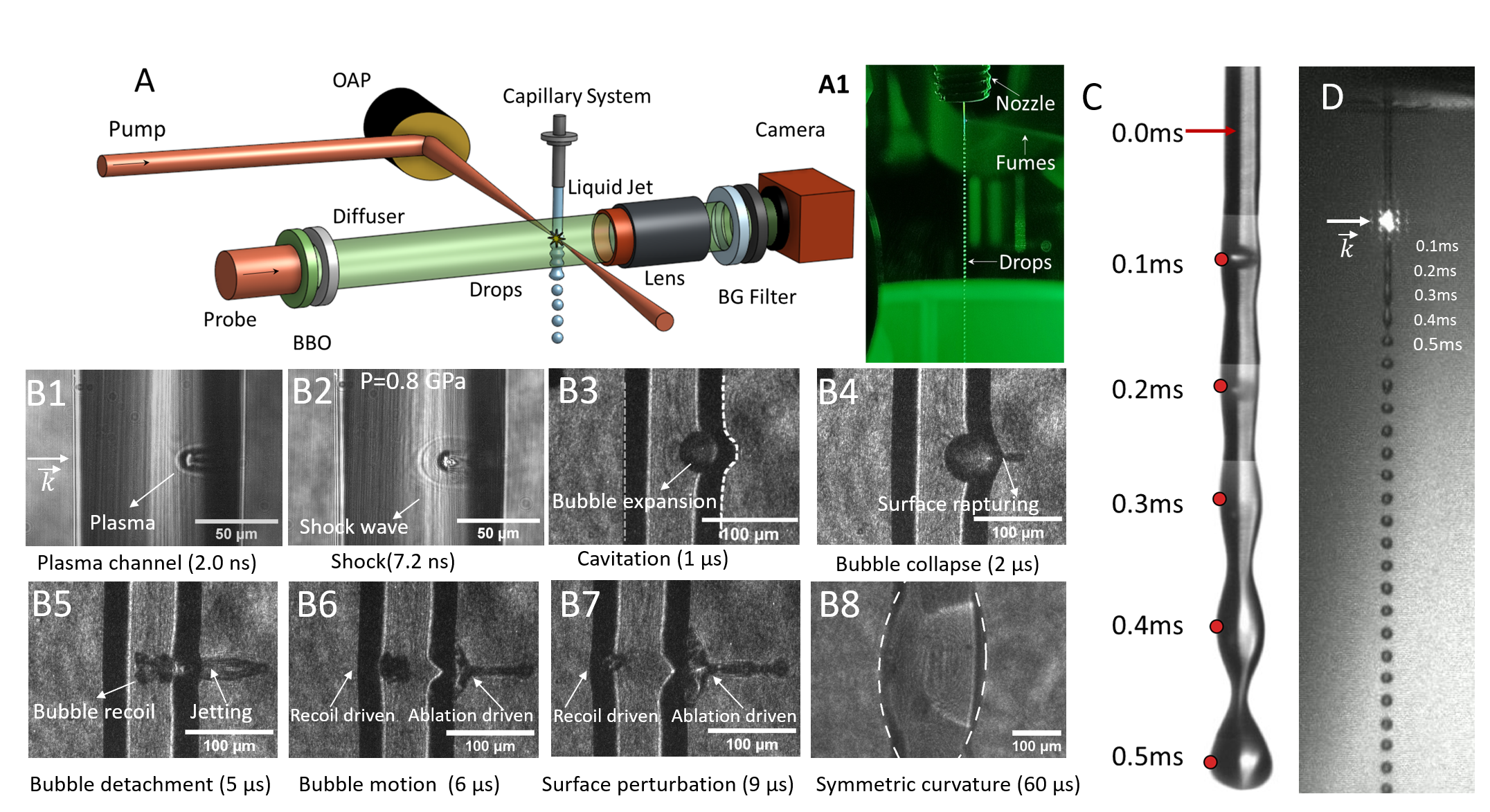}
\caption{\textbf{Generation and evolution of laser-seeded periodic perturbations.}
\textbf{A)} A multi-kHz femtosecond laser pulse (pump) is focused onto a laminar liquid jet to seed a localized surface perturbation, and the resulting dynamics are recorded with a time-delayed probe pulse \textbf{(A1)}. \textbf{(B1)} Localized optical breakdown followed by saturated plasma expansion at $1.8~\text{ns}$. \textbf{(B2)} A shock wave of estimated peak pressure ~0.8~GPa traverses the jet, with pulse width $\sim5~\mu\text{m}$ and velocity $v_s\approx2.0~\text{km/s}$, corresponding to a shock pulse duration of $\tau_{\text{shock}}\approx3~\text{ns}$; the full acoustic transit time across the $100~\mu$m jet is $\tau_{\rm ac}\approx70~\text{ns}$, both far shorter than the jet's capillary response time ($\tau_{\text{cap}}\sim42~\mu\text{s}$). \textbf{(B3)} A cavitation bubble nucleates near the rear interface and expands over microsecond timescales. \textbf{(B4)} Premature rupture at the rear free surface ($1$--$2~\mu\text{s}$) produces vapor ejection and recoil. \textbf{(B5 - B7)} The recoil wave traverses the jet and deforms the front surface, generating asymmetric perturbation. \textbf{(B8)} Over a timescale $\sim\tau_{\text{cap}}$, capillary relaxation suppresses the azimuthal asymmetry, yielding an axisymmetric mode suitable for Rayleigh--Plateau growth. \textbf{C)} The seeded mode advects downstream and amplifies exponentially, leading to necking and breakup (red arrow: seed location; red markers: successive laser pulses). \textbf{D)} Wide-field image showing periodic laser-induced perturbations and the resulting equidistant monodisperse droplet train.}
\label{fig1_dynamics}
\end{figure}

The first persistent surface deformation emerges on microsecond time scales through cavitation dynamics. The expanding cavitation bubble carries sufficient energy to overcome the capillary barrier for surface deformation and, in our geometry, ruptures the rear interface shortly after nucleation (Fig.~1B3--B4)~\cite{Energy_cavitation}. This rupture ejects a directional microjet into air~\cite{Patrascioiu2014,BlakeGibson1987} and, by conservation of momentum, drives a recoil flow across the jet column (Fig.~1B5) that deforms the front interface several microseconds later (Fig.~1B6--B7). 
Although optical breakdown, cavitation growth, rupture, and recoil are all strongly asymmetric processes, the small jet diameter ($D\sim100~\mu$m) relative to the capillary length scale allows the resulting disturbance to equilibrate azimuthally well within one capillary time, yielding a near-axisymmetric deformation (Fig.~1B8) (see Fig.~\ref{IP_growth_dependence}  for details). This axisymmetric seed then advects downstream and amplifies exponentially according to the Rayleigh--Plateau dispersion relation, leading to necking and breakup (Fig.~1C). Repetition of this sequence converts isolated laser-induced deformations into a train of perturbations convected downstream with the liquid flow (Fig.~1C,D)~\cite{Convective_advective_PRL}. This approach provides two independently controllable parameters: the perturbation spacing, set by the laser repetition rate, and the perturbation amplitude, set by the pulse energy.


\subsection*{Mode selection on the Rayleigh--Plateau dispersion curve}

We now test whether the laser-seeded perturbations follow the classical Rayleigh--Plateau dispersion relation~\cite{Plateau1857, Rayleigh1878, Rayleigh1892_minimal}. By varying the perturbation spacing, we sample different regions of the Rayleigh--Plateau spectrum; any long-lived flows or nonlocal deformations from cavitation would shift the selected wavenumbers away from this curve. To probe the instability spectrum, we vary the laser repetition rate $f_{\rm rep}$ at fixed pulse energy ($E\approx[1.8]~\mu$J, well below $E_c$; see Fig.~3). This directly tunes the perturbation spacing and hence the dimensionless wavenumber \(kR\), providing continuous access to both unstable and stable regions of the Rayleigh--Plateau spectrum. Unlike nozzle-driven experiments \cite{Lindblad1965, Dimmock1950}, where perturbations originate at the nozzle, perturbations here are seeded directly within the flowing column, enabling individual modes to be selected independently of upstream boundary conditions.

\begin{figure}[h!]
	\centering
	\includegraphics[width=1.0\textwidth]{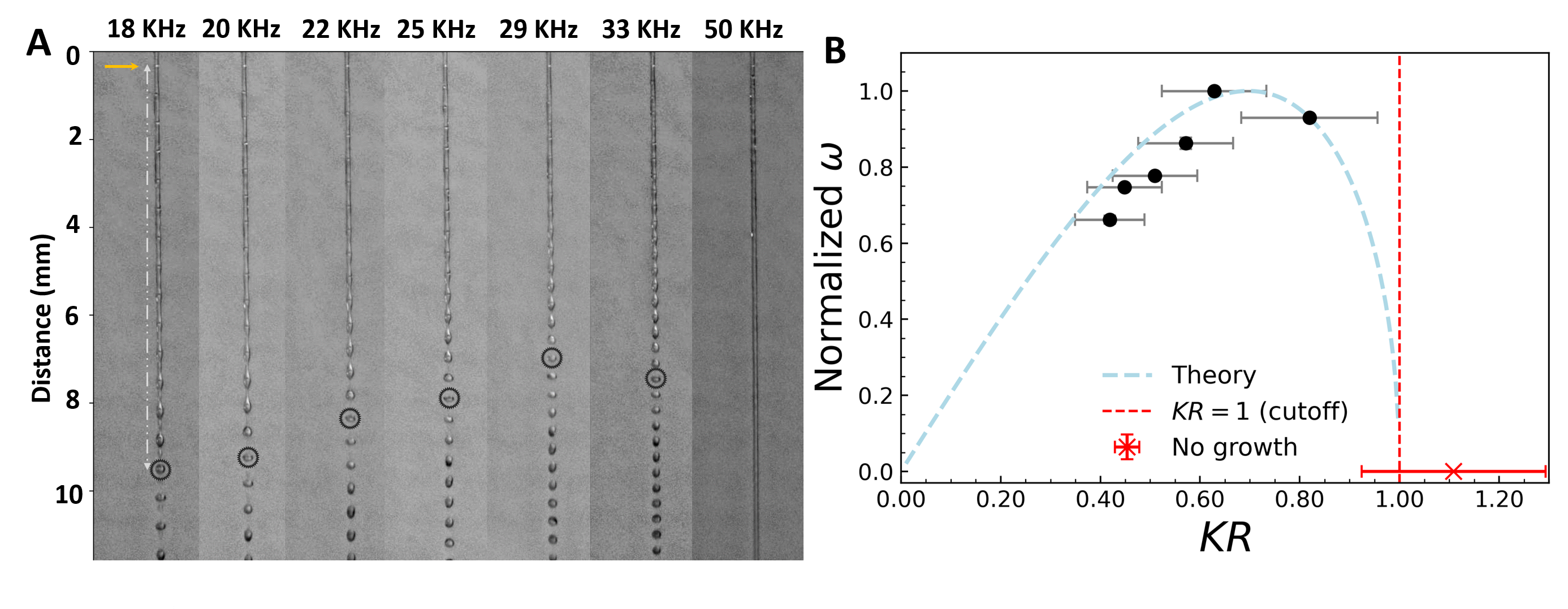}
	\caption{\textbf{Frequency-tuned optical seeding of Rayleigh--Plateau modes.} 
	\textbf{(A)} Representative images of the liquid jet perturbed by femtosecond laser pulses at varying repetition rates ($f_{\text{rep}}$). Orange arrow: laser interaction point; vertical dashed white arrow: breakup length; black circles: first droplet pinch-off location. At 18 kHz ($kR = 0.4$), perturbations grow slowly, yielding a long breakup length. As frequency increases toward 29--30 kHz, the breakup length reaches a minimum, indicating maximum growth rates. At 50 kHz ($kR = 1.1$), the jet remains intact; the perturbation lies outside the linearly unstable band ($kR>1$) and decays rather than growing, consistent with the stable branch of the inviscid Rayleigh--Plateau dispersion relation. \textbf{(B)} Normalized growth rate ($\omega$) versus $kR$. Experimental data agree with the classical Rayleigh--Plateau dispersion curve (dashed line), confirming the stability cutoff at $kR = 1$.}
	\label{fig:FM_seeding}
\end{figure}

Figure~2A shows the evolution of the breakup dynamics as a function of \(kR\). At long wavelengths (\(kR\ll1\)), perturbations grow slowly and breakup occurs only after extended propagation distances. As the repetition rate increases, the breakup length decreases and reaches a minimum near \(kR \approx 0.7\), consistent with the most unstable mode predicted by classical Rayleigh--Plateau theory \cite{Rayleigh1878, Plateau1857}. Further increases in repetition rate shift the perturbation into the short-wavelength regime. Near the stability boundary ($kR \approx 1$), perturbation growth becomes progressively weaker, consistent with $\omega\to0$ as $kR\to1$ in the classical dispersion relation~\cite{Rayleigh1878, Plateau1857}, while for $kR > 1$ the seeded disturbances lie on the stable branch, decay, and the jet recovers its unperturbed cylindrical profile..

We quantify the instability by measuring the spatial amplification of the seeded perturbations and extracting a normalized growth rate \(\omega\). As shown in Fig.~2B, the measured dispersion relation closely follows the classical Rayleigh--Plateau prediction, with maximum growth near \(kR\approx0.7\) and a stability cutoff at \(kR=1\). This confirms that, although the perturbations are seeded by cavitation-driven recoil following optical breakdown, they evolve according to the same instability spectrum as predicted for small-amplitude capillary disturbances. By tuning the laser repetition rate, we can therefore select Rayleigh--Plateau modes directly within the flowing column, providing a contact-free alternative to conventional nozzle-driven jets.


\subsection*{Amplitude-driven transition from linear to nonlinear Rayleigh--Plateau seeding}

We now examine how the amplitude of the laser-seeded perturbation controls the crossover from linear, single-mode Rayleigh--Plateau growth to strongly nonlinear, broadband seeding~\cite{Yuen1968,Crane1964,Reversal2022,Nayfeh1970,Chaudhary1980b, Chaudhary1980c, Eggers_RPP2008}. To isolate this effect, we fix the excitation wavelength near the maximum of the Rayleigh--Plateau growth curve (experimentally accessible point $kR\approx0.65$, close to the theoretical maximum $kR\approx0.697$) and vary the laser pulse energy ($E$). This enables a direct comparison with classical linear theory at low amplitude and reveals the onset of nonlinear seeding at higher energies (see~Fig.~\ref{energy_perturbation_strength} for the perturbation strength induced by bubble recoil as a function of laser energy).

Figure~3A displays the evolution of the jet breakup length. Without optical seeding, spontaneous breakup occurs via ambient noise amplification. With increasing pulse energy, breakup shifts progressively closer to the interaction point (Supplementary Fig.~\ref{breakup_reversal}), and the measured breakup length $L_b$ decreases linearly with $\ln E$ over the range $E=[1.2]$--$[2.2]~\mu$J (Fig.~3B) This behavior is modeled by classical linear theory ($\delta_0 \ll R$), where the surface deformation scales as $\delta(z) = \delta_0 \exp(\omega z / U)$ \cite{Rayleigh1878, Rayleigh1892_minimal,Eggers_RPP2008}, with $U$ being the jet velocity and $\omega$ the growth rate. Applying the terminal pinch-off condition $\delta(L_b) \sim R$ yields $L_b = (U/\omega) \ln(R/\delta_0)$.  Because $\omega$ is strictly constant at a fixed $kR$, this log-linear form holds for any power-law relation $\delta_0\propto E^\alpha$, so the observed scaling is consistent with pure Rayleigh--Plateau growth from a seed whose amplitude increases monotonically with pulse energy, independent of the specific form of $\delta_0(E)$. This is corroborated by the direct amplitude calibration in Supplementary Fig.~\ref{energy_perturbation_strength}, confirming that the laser systematically sets $\delta_0$.

\begin{figure}[h!] 
	\centering
	\includegraphics[width=1.0\textwidth]{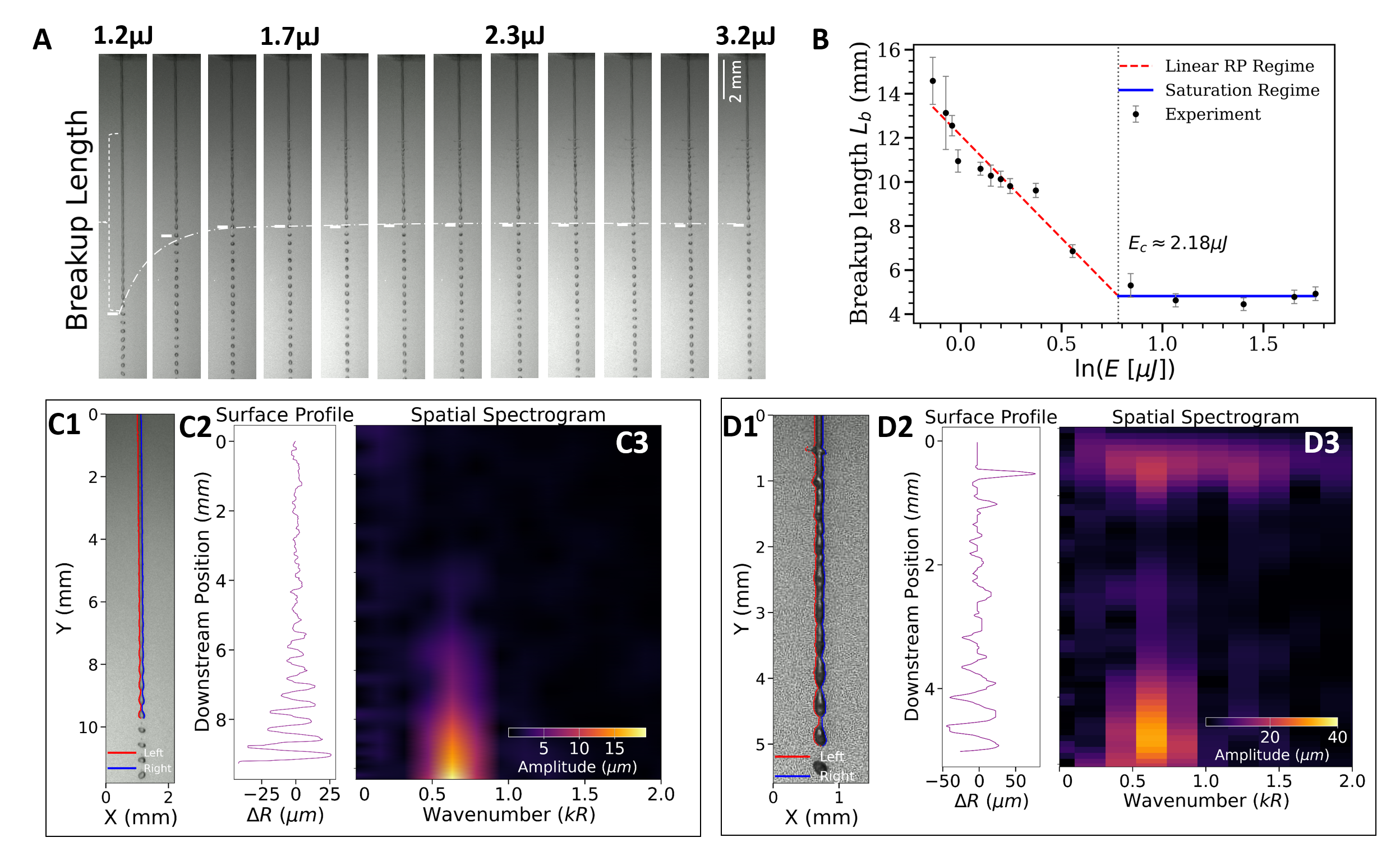}
	\caption{\textbf{Amplitude control of seeded surface perturbations and breakup:} \textbf{(A)} Raw images of jet breakup at different pulse energies near the resonance frequency (29 kHz). In the absence of laser forcing (0 $\mu$J), the jet undergoes ambient noise-driven breakup. With increasing pulse energy, the breakup length decreases, eventually reaching a saturation plateau. \textbf{(B)} Experimental breakup length $L_b$ versus $\ln E$. The red dashed line is a linear–log fit ($L_b = \mathrm{const} - B\ln E$), as expected from linear Rayleigh–Plateau amplification of a seed whose amplitude increases monotonically with pulse energy. A transition occurs at $E_c = 2.18~\mu$J, identified as the intersection of the log-linear fit and the saturation-plateau fit (vertical dotted line); beyond this, $L_b$ enters a saturation plateau (blue solid line), consistent with the perturbation amplitude approaching the jet radius ($\epsilon\sim R$), i.e., strongly nonlinear seeding.. \textbf{(C, D)} Spatial spectrograms of surface perturbations across the transition from linear growth to nonlinear evolution, with a sharp crossover at $E_c$ where the breakup length saturates. (C) Linear-growth regime ($E < E_c$); (D) saturation regime ($E > E_c$). Sub-panels (C1, D1) show the detected jet boundaries extracted from the images. \textbf{(C2, D2)} plot the downstream evolution of liquid jet radius perturbations. \textbf{(C3, D3)} display the corresponding spatial spectrograms of the dimensionless wavenumber $kR$. In the linear regime (C3), the spectrum shows amplification of the laser-driven mode, which lies near the most unstable Rayleigh--Plateau mode ($kR \sim 0.65$). In the saturation regime (D3), the large-amplitude seed contains a broad spectrum of modes; along the jet, this laser-driven mode grows, while higher-wavenumber modes ($kR > 1.0$) decay, consistent with the Rayleigh--Plateau stability criterion.}
    
\label{Shadow Image}
\end{figure}

Beyond this point, the log-linear scaling breaks down, with $L_b$ departing from the logarithmic decrease at $E_c \approx 2.18~\mu\text{J}$ and reaching a saturation plateau(Fig.~3B). Measurements of breakup length and surface perturbation growth alone do not reveal the origin of this saturation; we therefore analyze the spatial evolution of the perturbation spectrum below and above $E_c$. In the weak-forcing regime ($E < E_c$), the initial perturbation just downstream of the laser focus is comparable to background surface fluctuations (Figs.~3C1, C2). Because the laser-imposed wavelength corresponds to $kR$ near the most unstable Rayleigh--Plateau mode, this mode experiences the largest growth rate and dominates over background fluctuations, as evidenced by the narrow spectral peak at $kR \approx 0.65$ in the spatial spectrogram (Fig.~3C3).

In contrast, for $E > E_c$ the jet enters a qualitatively different regime. The laser impact produces strong ablation and cavitation-driven jetting, creating a large-amplitude, strongly distorted surface deformation at the interaction point (Figs.~3D1, D2). This non-sinusoidal profile contains significant higher-order spatial harmonics, as seen in the spatial spectrogram (Fig.~3D3), which shows broadband excitation extending well beyond the harmonics of the driven fundamental alone. This is consistent with the sharply localized, high-curvature deformation left by cavitation-driven interface rupture and jetting (Fig.~1B4), whose broadband spectral content reflects the spatial sharpness of the initial disturbance rather than a distributed periodic forcing. Downstream advection filters this spectrum, with modes having $kR > 1$ decaying rapidly while only modes in the classical unstable band ($kR < 1$) grow, dominated by the driven mode at $kR \approx 0.65$. The critical energy $E_c$ thus marks the transition from linear, single-mode amplification to a regime where strong nonlinear deformation spreads energy over many modes, most of which do not significantly contribute to breakup.


\subsection*{Nonlinear breakup beyond the Rayleigh--Plateau stability limit}

Above the critical energy $E_c\approx2.18~\mu$J measured at $kR\approx0.65$ (Fig.~3B), the perturbation dynamics depart from the single-mode picture of linear Rayleigh--Plateau growth. This raises the question of whether sufficiently strong forcing can drive breakup at $kR>1$, where infinitesimal perturbations are linearly stable. We find that sufficiently strong, sustained local perturbations indeed drive deterministic droplet pinch-off in this linearly stable regime. To probe the regime beyond the classical stability limit, we increase the repetition rate to excite perturbations at $kR\approx1.8$--$1.9$. Under weak forcing ($E<2~\mu$J), the laser-induced perturbation drives transient necking down to a minimum radius of $18.6~\mu$m ($\approx0.37R$), before relaxing back toward the unperturbed jet radius. The FFT spectrum confirms excitation at $kR>1$, where the Rayleigh--Plateau instability does not amplify perturbations. They instead remain on the linearly stable branch of the Rayleigh--Plateau dispersion relation, and the modal content decays and redistributes among multiple wavenumbers, as in Fig.~3D.

\begin{figure}[h!]
    \centering
    \includegraphics[width=0.9\textwidth]{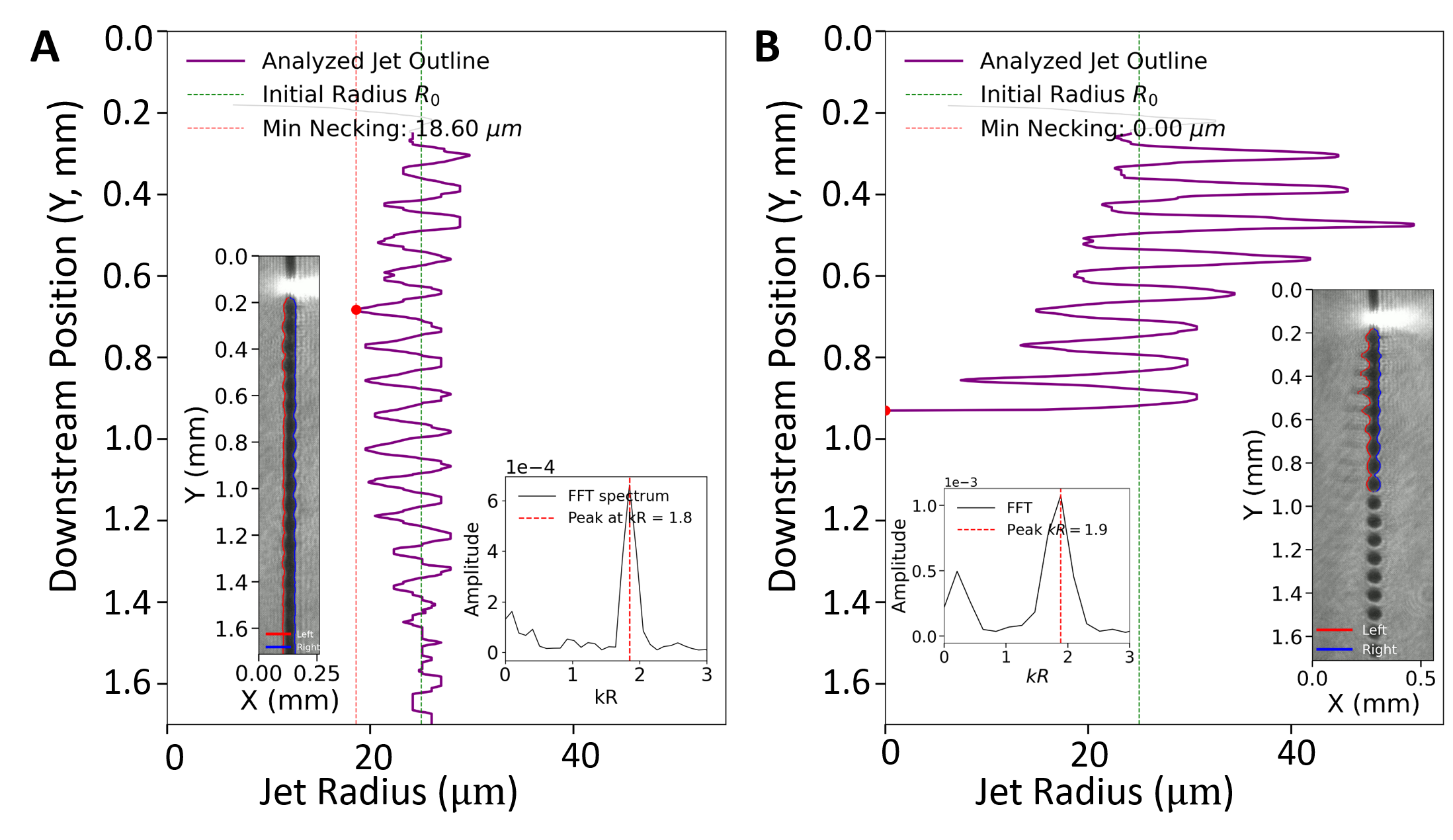}
    \caption{\textbf{Breaking the Rayleigh–Plateau stability limit via strong optical perturbations.} Measurements in this figure use a jet of diameter $\approx50~\mu$m ($R\approx25~\mu$m), smaller than the $R=50~\mu$m jet used in Figs.~2--3; $kR$ values are computed from this jet's own measured radius and velocity. \textbf{(A)} Evolution of surface modulations at low pulse energy ($E < 2\mu$J) for $kR = 1.8$. 
    (\textbf{A}) Evolution of surface modulations at low pulse energy ($E < 2\mu$J) for $kR = 1.8$. The FFT inset (top) confirms the excitation of a stable mode. The main panel tracks the jet contour; although transient necking reduces the radius to a minimum of 18.6~$\mu$m (red dashed), the perturbation eventually decays and the jet returns to its laminar state (green dashed). \textbf{B}) High-energy regime ($E > 3~\mu$J) at $kR = 1.9$. The FFT inset shows the dominant mode. Here, sustained cavitation-driven jetting and recoil (Section~1, Fig.~1B4--B7) prolongs the necking beyond the point of recovery, driving the neck radius to zero and resulting in droplet pinch-off at a rate of $0.2$~MHz. This demonstrates breakup in a domain where linear theory predicts stability, driven by sufficiently strong and sustained localized forcing rather than by linear modal growth.} 
    \label{fig:placeholder}
\end{figure}

Above this threshold ($E > E_c$), the evolution of the seeded perturbation for $kR>1$ transitions from Rayleigh--Plateau-dominated decay to a cavitation-driven nonlinear necking regime.  For $E>3~\mu$J, each laser pulse induces sustained cavitation-driven jetting and interface deformation that persists for $\sim35~\mu$s, more than twice the capillary time of this jet ($\tau_{\rm cap}\approx14.9~\mu$s, $R=25~\mu$m; Supplementary Section~1.4) In contrast to the transient, recovering necking seen at lower energy (Fig.~4A), the neck radius here decreases continuously and irreversibly to pinch-off, despite excitation at a linearly stable wavenumber of $kR=1.9$ (Fig.~4B). This irreversibility follows because the forcing duration ($\sim35~\mu$s) exceeds the local capillary relaxation time, preventing the recoil that restores the jet to its unperturbed profile at lower energy: breakup at $kR>1$ is therefore governed jointly by the perturbation amplitude and its duration relative to $\tau_c$, not by amplitude alone. 

At a fixed repetition rate of $40~\mathrm{kHz}$, increasing the pulse energy from $1.8~\mu\mathrm{J}$ to $2.4~\mu\mathrm{J}$ changes the response from no breakup to droplet formation, further demonstrating that increasing the perturbation amplitude extends the accessible breakup regime beyond the classical $kR=1$ cutoff (see Fig.~\ref{stronglyperturbed}D in the Supplementary Information). This establishes localized femtosecond-laser cavitation as a means of driving deterministic jet breakup beyond the classical Rayleigh--Plateau stability limit.


\begin{figure}[h!]
	\centering
	\includegraphics[width=1.0\textwidth]{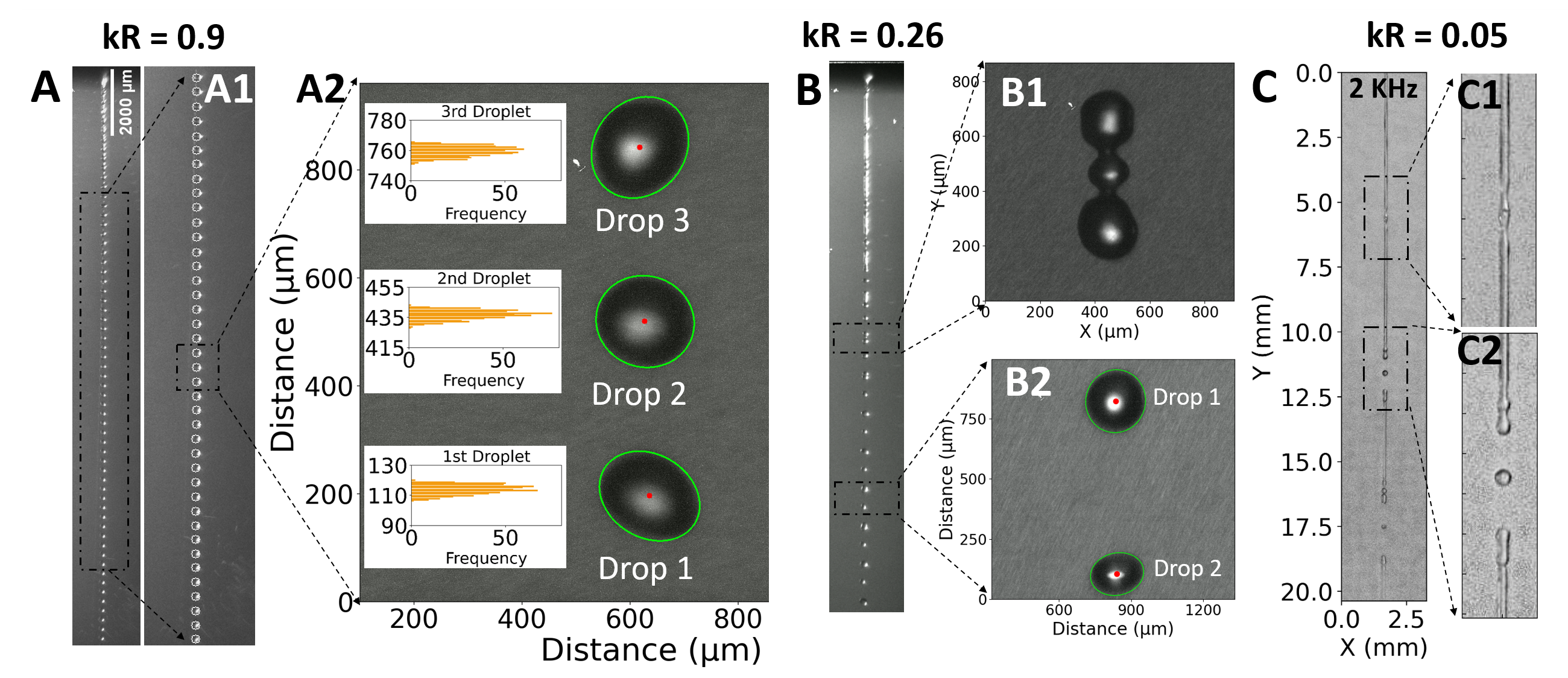} 
	\caption{\textbf{Wavenumber-controlled breakup: from synchronized droplet trains to isolated droplets.} 
	\textbf{A}) Broad-field shadowgraph of a laser-perturbed jet at $kR=0.9$, near the maximum Rayleigh–Plateau growth rate, producing a synchronized droplet train.
	\textbf{A1}) Downstream magnification of \textbf{A}), showing long-range droplet uniformity.
	\textbf{A2}) Representative droplets with extracted contours (green line) and centroids (red point); accompanying histograms quantify positional stability.
	\textbf{B}) At $kR=0.26$, the longer perturbation wavelength produces elongated liquid columns ($\Lambda = L/D \sim 10$) that undergo capillary contraction. 
	\textbf{B1}) Dumbbell-shaped intermediate formed during capillary collapse. \textbf{B2})Two droplet populations emerge: a larger droplet from capillary retraction (\textbf{B1}) and a smaller droplet from the primary laser-seeded pinch-off. 
	\textbf{C}) Surface deformation at the extreme long-wavelength limit ($kR=0.05$). 
	\textbf{C1}) At $kR=0.05$, the large inter-pulse spacing ($\sim6000,\mu$m, $\sim60D$) suppresses coupling between neighboring perturbations, causing each laser-seeded excitation to evolve as an isolated hydrodynamic wave packet. 
	\textbf{C2}) The isolated perturbation evolves as a localized wave packet, producing single-droplet pinch-off. The remaining liquid column ($\Lambda>60$) subsequently fragments through capillary instability despite  the liquid having low Ohnesorge number ($Oh$).}
	\label{kr_scan_localized}
\end{figure}

\subsection*{Wavenumber-controlled transition from synchronized droplet trains to isolated droplets}

By exploiting the localized cavitation-driven seed, we traverse the unstable Rayleigh--Plateau band from $kR = 0.9$ to the extreme long-wavelength regime $kR = 0.05$, where synchronized droplet trains give way to isolated primary-droplet formation (Fig.~5). Unlike nozzle-based excitation, which fixes the disturbance pattern at the jet origin, the localized seed acts directly on the evolving liquid column, allowing successive perturbations to develop with weak or strong mutual coupling depending on their separation. 

At $kR=0.9$, within the Rayleigh--Plateau unstable band, periodic optical forcing produces a synchronized monodisperse droplet train (Fig.~5A). The corresponding wavelength is $\lambda/D\simeq3.5$, and the imposed spacing is preserved during downstream growth and breakup. High-magnification imaging of 40 successive droplets recorded over $\sim47$~s (Fig.~5A1--A2) shows highly reproducible droplet morphology, with major and minor axes of $183\pm2~\mu\mathrm{m}$ and $169\pm1~\mu\mathrm{m}$, respectively — a shape variation of about 1$\%$ (Supplementary Fig.~\ref{dropletstability}E--F). Droplet centroid positions are stable to within the imaging resolution ($\lesssim11~\mu\mathrm{m}$, comparable to one camera pixel).

Reducing the imposed wavenumber to $kR=0.26$ increases the perturbation wavelength~\cite{PadillaMartinez2013}  to $\lambda/D\simeq10$ (Fig.~5B). The resulting elongated liquid segments first contract under surface tension, forming transient dumbbell-like structures (Fig.~5B1), before fragmenting into two droplet populations (Fig.~5B2)~\cite{Taub1976}. The smaller droplets result from the primary laser-seeded pinch-off, whereas the larger droplets form through capillary retraction of the remaining liquid segment. Thus, increasing the perturbation wavelength introduces a secondary collapse process and changes the breakup from monodisperse to bidisperse. The temporal and morphological stability of the resulting droplet trains, including their positional jitter and shape fluctuations, is analyzed in the Supplementary Information (Figs.~\ref{dropletstability} and ~\ref{localizedperturbation}).

At the extreme long-wavelength limit, $kR=0.05$, the perturbation spacing reaches $\lambda\simeq6000~\mu\mathrm{m}\simeq60D$ (Fig.~5C). The large separation between successive seeds reduces their mutual interaction, allowing each laser-induced deformation to evolve locally. Each deformation produces one primary droplet by capillary pinch-off, while the residual liquid column, extending over more than $60D$, subsequently undergoes secondary breakup (Fig.~5C1--C2). This sequence separates the laser-triggered pinch-off from the later fragmentation of the remaining column, enabling isolated primary-droplet formation ~\cite{BardeauIsolatedDrops,Garcia_PRE2019,Zhang_PRE2026}. The secondary breakup proceeds via the classical inertio-capillary Rayleigh--Plateau cascade expected at this low Ohnesorge number ($Oh\approx0.01$ for the $R=50~\mu$m jet; Supplementary Section~1.4), for which viscous effects do not suppress the instability. These wavelength-dependent breakup states define the low-$kR$ branch of the two-parameter dynamical map presented in Fig.~6.


\subsection*{Wavenumber--amplitude map of Rayleigh--Plateau breakup and beyond}

The laser repetition rate and pulse energy independently control the perturbation wavenumber $kR$ and amplitude $\epsilon/R$, respectively, defining a two-dimensional parameter space in which the distinct breakup regimes observed above are organized (Fig.~6). The classical Rayleigh--Plateau dispersion relation provides the reference structure, with its maximum near $kR\simeq0.69$ and cutoff at $kR=1$.

Below the optical-breakdown threshold, $E<E_{\mathrm{th}}\simeq1.2~\mu\mathrm{J}$, no cavitation-driven seed forms and breakup remains background-driven and stochastic. For $E>E_{\mathrm{th}}$, the laser seed dominates the background and the response becomes deterministic. In the central unstable region, approximately $0.4\lesssim kR\lesssim0.9$, weak-to-moderate forcing produces synchronized monodisperse droplets, with the breakup length decreasing approximately linearly with $\ln E$. Increasing the forcing beyond $E_{\mathrm{c}}\simeq2.18~\mu\mathrm{J}$ (measured at $kR\approx0.65$) produces the broadband nonlinear regime identified in Fig.~3, where the breakup length saturates as the initial deformation becomes strongly non-sinusoidal.

\begin{figure}[h!]
	\centering
	\includegraphics[width=0.975\textwidth]{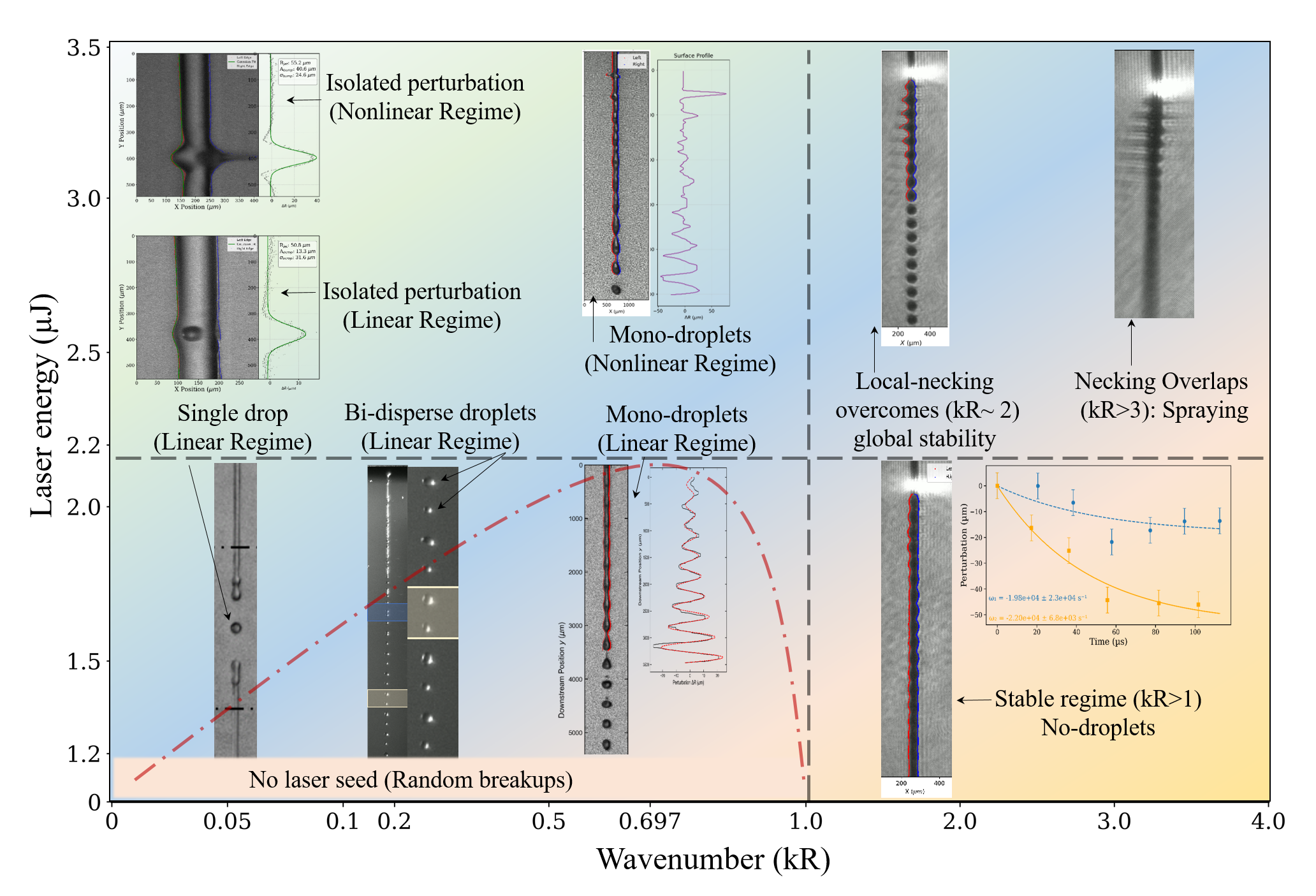} 
	  \caption{\textbf{Laser-driven perturbation landscape.}
    Horizontal axis: dimensionless wavenumber $kR$, controlled via laser repetition rate. Vertical axis: seeded perturbation ratio $\epsilon/R$, tuned via laser energy $E$). The landscape, classified by distinct dynamics, is guided by the Rayleigh--Plateau dispersion relation (dashed curve, max growth at $kR \approx 0.69$, cutoff at $kR = 1$). \textbf{(i)} Threshold regime ($E < E_{\text{th}} \approx 1.2~\mu\text{J}$): laser-seeded perturbation fails to overcome the convected upstream noise on the liquid jet; breakup is random.	\textbf{(ii)} Linear single-mode ($E_{\text{th}} < E < E_{\text{c}}$, $kR \approx 0.4$--$0.9$): monodisperse droplets, breakup length scales log-linearly with $E$. 	\textbf{(iii)} Nonlinear broadband ($E > E_{\text{c}}$, central $kR$): the seed spectrum broadens to include stable high-$k$ modes, which decay downstream; breakup length saturates. \textbf{(iv)} Bi-disperse regime (low $kR \approx 0.2$--$0.3$): local pinching and long-column collapse produce two droplet populations. \textbf{(v)} Isolated perturbation ($kR \approx 0.05$): decoupled wave packets generate single isolated droplets with $L/D > 60$ segments. \textbf{(vi)} Stable regime ($kR > 1$, low $E$): perturbations decay; no breakup. 	\textbf{(vii)} Stability-breaking ($kR > 1$, $E > E_{\text{c}}$): sustained cavitation-driven jetting and recoil enable pinch-off up to $kR \approx 1.9$. \textbf{(viii)} Overlap/spray limit ($kR > \pi$): successive perturbations overlap, preventing necking; induces spraying. This phase-map establishes localized optical forcing as a continuously tunable platform spanning linear instability, finite-amplitude dynamics, and isolated hydrodynamic excitations.}
	\label{fig:kr_phase_space_scan}
\end{figure}

At low $kR$, the map contains the wavelength-controlled sequence established in Fig.~5: synchronized monodisperse trains near $kR\simeq0.9$, bidisperse breakup at $kR\simeq0.2$--$0.3$, and isolated primary-droplet formation near $kR\simeq0.05$, where successive seeds are separated by $\lambda/D\simeq60$. At $kR=0.05$, the isolated-seed response depends strongly on pulse energy. For weak forcing ($E\sim2$--$3~\mu\mathrm{J}$), the deformation is broad, with $A/R=0.26$ and axial radius of curvature $R_z\simeq80~\mu\mathrm{m}$, larger than the jet radius. At higher energy ($E\sim10$--$20~\mu\mathrm{J}$), the deformation becomes strongly nonlinear and sharply localized, with $A/R=0.74$ and $R_z\simeq16~\mu\mathrm{m}$. Thus, increasing the pulse energy changes not only the perturbation amplitude but also the relative contributions of the axial and azimuthal curvatures to the capillary response. The low-$kR$ branch therefore provides access to both weakly deformed isolated seeds and strongly localized finite-amplitude perturbations.

Beyond the classical cutoff, the response is strongly amplitude dependent. For $kR>1$ and weak forcing, the imposed perturbations decay without pinch-off. At sufficiently large amplitude, localized cavitation-driven forcing drives irreversible necking to capillary pinch-off at $kR\simeq1.9$, despite the linear stability of the corresponding mode. For $kR\gtrsim\pi$, the seed spacing becomes comparable to or smaller than the jet diameter, $\lambda/D\lesssim1$. Successive cavitation-induced deformations then overlap before distinct valleys can develop, suppressing independent neck formation and driving the response toward irregular spraying rather than discrete droplet pinch-off.

The $(kR,\epsilon/R)$ map shows how the two control parameters determine the breakup outcome. Changing $kR$ changes the spacing of the imposed perturbations, producing synchronized droplet trains, bidisperse breakup, isolated droplets (see Fig.~\ref{local_to_multimhz}), or decay in the $kR>1$ regime (see Fig.~\ref{stronglyperturbed}). At a fixed $kR$, increasing $\epsilon/R$ changes the response from selective Rayleigh--Plateau growth to broadband nonlinear deformation and, for sufficiently strong forcing, irreversible pinch-off even when $kR>1$ (see Fig. \ref{stronglyperturbed}). Thus, localized femtosecond seeding controls both the spacing and amplitude of the initial disturbance while leaving the classical dispersion relation unchanged.

\section*{Conclusion}
We have used localized femtosecond-laser-induced cavitation to seed and control Rayleigh--Plateau breakup in a freely flowing liquid jet. The cavitation recoil provides a localized, finite-amplitude disturbance whose spacing and strength are set independently by the laser repetition rate and pulse energy. At low energies, the imposed perturbations recover the classical Rayleigh--Plateau dispersion relation and its $kR=1$ cutoff. Increasing the pulse energy instead drives nonlinear evolution, culminating in broadband seeding and finite-amplitude pinch-off; independently, varying the perturbation wavelength moves the jet from synchronized monodisperse trains to bidisperse breakup and, at the longest wavelengths, isolated primary droplets.

The resulting $(kR,\epsilon/R)$ map therefore organizes linear instability, nonlinear multimode evolution, and finite-amplitude pinch-off—including deterministic breakup for $kR>1$—within a single experimental platform. In particular, strong localized forcing produces deterministic breakup up to $kR\simeq1.9$, where infinitesimal perturbations are linearly stable. The laser therefore controls the initial disturbance without altering the classical dispersion relation, enabling deterministic jet breakup across and beyond the Rayleigh--Plateau stability band, with laser-synchronized primary-droplet generation reaching $0.2~\mathrm{MHz}$. Together, these results establish localized femtosecond-laser-induced cavitation as an experimental framework for studying Rayleigh--Plateau dynamics from linear instability to finite-amplitude pinch-off, while providing a route to controlled, high-rate droplet generation.


\bibliographystyle{naturemag}
\bibliography{references}



\section*{Methods}

\subsection*{Laser parameters and liquid-jet target}

Experiments were performed at TIFR-Hyderabad, India, using a commercial Yb:KGW regenerative amplifier laser system (Pharos, Light Conversion Ltd.) delivering $\sim$175 fs pulses at a central wavelength of 1030 nm, with a tunable repetition rate up to 200 kHz and pulse energies up to 2000 $\mu$J. The 7 mm diameter laser beam propagated horizontally, perpendicular to the liquid-jet axis, and was focused using a gold-coated off-axis parabolic mirror ($f=100.16$ mm). The focal spot was characterized using a $20\times$ Mitutoyo NIR objective (NA = 0.42), yielding a $1/e^2$ beam waist of $12.9~\mu$m $\times$ $11.4~\mu$m. For a pulse energy of $1~\mu$J, the corresponding vacuum peak intensity at focus was estimated to be $1.27\times10^{12}~\mathrm{W\,cm^{-2}}$. Upon interaction with the transparent cylindrical liquid jet, however, cylindrical lensing enhanced the local irradiation beyond $10^{13}~\mathrm{W\,cm^{-2}}$, exceeding the optical breakdown threshold of water and initiating localized plasma formation at the rear interface of the jet.

The target consisted of a downward-flowing deionized water jet operated under ambient conditions ($25^\circ$C, 1 atm). The jet was produced using a home-built metallic nozzle fabricated by laser micromachining. A nitrogen-backed pressure of 2--4 bar generated stable laminar jets with diameters ranging from 90 to 130 $\mu$m (typically $\sim100~\mu$m) and flow velocities of 10--20 m\,s$^{-1}$. In the absence of laser irradiation, the jet exhibited a natural Rayleigh--Plateau breakup length of approximately 15 mm. To ensure that laser-seeded perturbations evolved independently of the natural breakup dynamics, the interaction region was positioned 2--3 mm below the nozzle exit. The nozzle position was adjusted using manual three-axis translation stages.

\subsection*{Nanosecond pump--probe shadowgraphy}

Early-time plasma and shock-wave dynamics were recorded using pump--probe shadowgraphy with an 8-bit CMOS camera. A fraction of the pump beam was frequency doubled in a $\beta$-barium borate (BBO) crystal to generate a 515 nm probe pulse. A diffuser was inserted into the probe beam path to provide spatially uniform illumination. Temporal delays in the nanosecond regime were introduced using an optical delay line. To suppress plasma self-emission and reject scattered fundamental pump light, a BG39 glass filter was placed in front of the imaging sensor. This configuration enabled visualization of plasma formation and the propagation of gigapascal-scale shock waves within the liquid jet.

\subsection*{Microsecond-to-millisecond fluid-dynamics imaging}

Longer-timescale fluid dynamics, including cavitation expansion, surface deformation, jetting, and perturbation evolution, were recorded using an intensified CCD (ICCD) camera coupled to the same $20\times$ Mitutoyo NIR objective. Illumination was provided by a continuous-wave red diode laser ($\lambda \approx 630$--680 nm). The ICCD was synchronized to the femtosecond pump pulse using a photodiode trigger and a digital delay generator, providing electronically adjustable delays from $1~\mu$s to $100~\mu$s. Gate widths between 4 and 50 ns were employed depending on the timescale of interest. A BG39 filter was again used to suppress residual plasma emission and scattered pump light. For tracking the long-term growth of laser-seeded perturbations, the same imaging system was used to record the spatial evolution of the jet along the flow direction; downstream position was converted to an effective evolution time using the measured jet velocity $U$ ($t = z/U$), corresponding to evolution times of approximately 0.04--0.5 ms. This approach enabled direct visualization of the transition from localized perturbations to necking, pinch-off, and the formation of an equidistant train of monodisperse droplets.

\section*{Acknowledgments}
\paragraph*{Funding:}
The authors acknowledge support of the Department of Atomic Energy, Government of India, under Project Identification No. RTI4007.
\paragraph*{Author contributions:} P.K.S. and S.S. conceived the idea and initiated the project. All the authors contributed in setting up the experiment. P.K.S. and S.S. wrote the manuscript. S.S. led the data analysis with input from other authors.
\paragraph*{Competing interests:}
The authors declare no competing interests.
\paragraph*{Data and materials availability:}
All data needed for evaluations of conclusions in the paper are mentioned in the main and/or in the Supplementary sections. Any additional data related to this paper can be requested from the authors.

\newpage

\renewcommand{\thefigure}{S\arabic{figure}}
\renewcommand{\thetable}{S\arabic{table}}
\renewcommand{\theequation}{S\arabic{equation}}
\renewcommand{\thepage}{S\arabic{page}}
\setcounter{figure}{0}
\setcounter{table}{0}
\setcounter{equation}{0}
\setcounter{page}{1} 


\clearpage
\begin{center}
{\LARGE\bfseries Supplementary Information}

\vspace{0.8cm}

{\Large\bfseries \scititle}

\vspace{0.8cm}

Sourabh Singh$^{1}$,
Tamanna$^{1}$,
S.~Sree~Harsha$^{1}$,
Prashant Kumar Singh$^{1,*}$

\vspace{0.4cm}

{\small
$^{1}$Tata Institute of Fundamental Research Hyderabad\\
36/P Gopanpally, Hyderabad 500046, Telangana, India
}

\vspace{0.3cm}

{\small
*Correspondence: \texttt{pks@tifrh.res.in}
}

\end{center}
\vspace{1cm}
\setcounter{section}{0}
\section*{Supplementary Text}

\section{From Optical Breakdown to Rayleigh--Plateau Seed Formation}

This section details the physical processes that convert a focused femtosecond laser pulse into a well-defined, axisymmetric Rayleigh--Plateau (RP) seed perturbation on a flowing liquid jet. It provides quantitative time-scale and energy arguments that establish the shock stage as dynamically subdominant, and identify cavitation-driven recoil as the dominant seeding mechanism.

\subsection{Experimental geometry and optical focusing}

The liquid jet is generated from a laminar nozzle with radius $R_{\mathrm{jet}} = 50~\mu\mathrm{m}$ ($D = 100~\mu\mathrm{m}$), the default geometry used throughout this work unless otherwise stated (see Section~4 for measurements using a smaller-diameter jet). The femtosecond laser pulse (wavelength, pulse duration, and repetition rate as specified in Methods) is focused into the jet using a microscope objective, yielding a $1/e^2$ beam-waist radius at the focus of $\omega_0 \approx 12~\mu\mathrm{m} < R_{\mathrm{jet}}$. This tight focusing ensures that the energy deposition is spatially localized on a scale much smaller than the jet radius.

The cylindrical jet acts as a strong refractive lens. For a water jet in air, the refractive index contrast ($n_{\mathrm{water}} \approx 1.33$, $n_{\mathrm{air}} \approx 1$) causes the incident beam to be concentrated near the rear liquid--air interface, forming a caustic region where the local intensity is significantly enhanced \cite{Caustic_Book, Cylinder_focusOL}. Ray-tracing calculations and prior measurements on similar geometries indicate intensity enhancement factors of order 5--10 at the rear interface, depending on the exact incidence angle and beam profile. This refractive focusing further localizes the optical breakdown to a micron-scale volume adjacent to the rear surface. At the input pulse energies used in these experiments ($E_{\rm in} \sim 2~\mu\text{J}$), and accounting for transmission losses and focusing geometry, the local peak intensity in the caustic region exceeds $I_{\rm peak} \gtrsim 10^{13}~\mathrm{W\,cm^{-2}}$, which is well above the threshold for optical breakdown in water via multiphoton, tunneling, and avalanche ionization processes~\cite{Optical_breakdown_Vogel, Optical_breakdown_Water}.

\subsection{Enhanced energy coupling via strong focusing by a liquid microlens}

\begin{figure}[h!]
\centering
\includegraphics[width=1\textwidth]{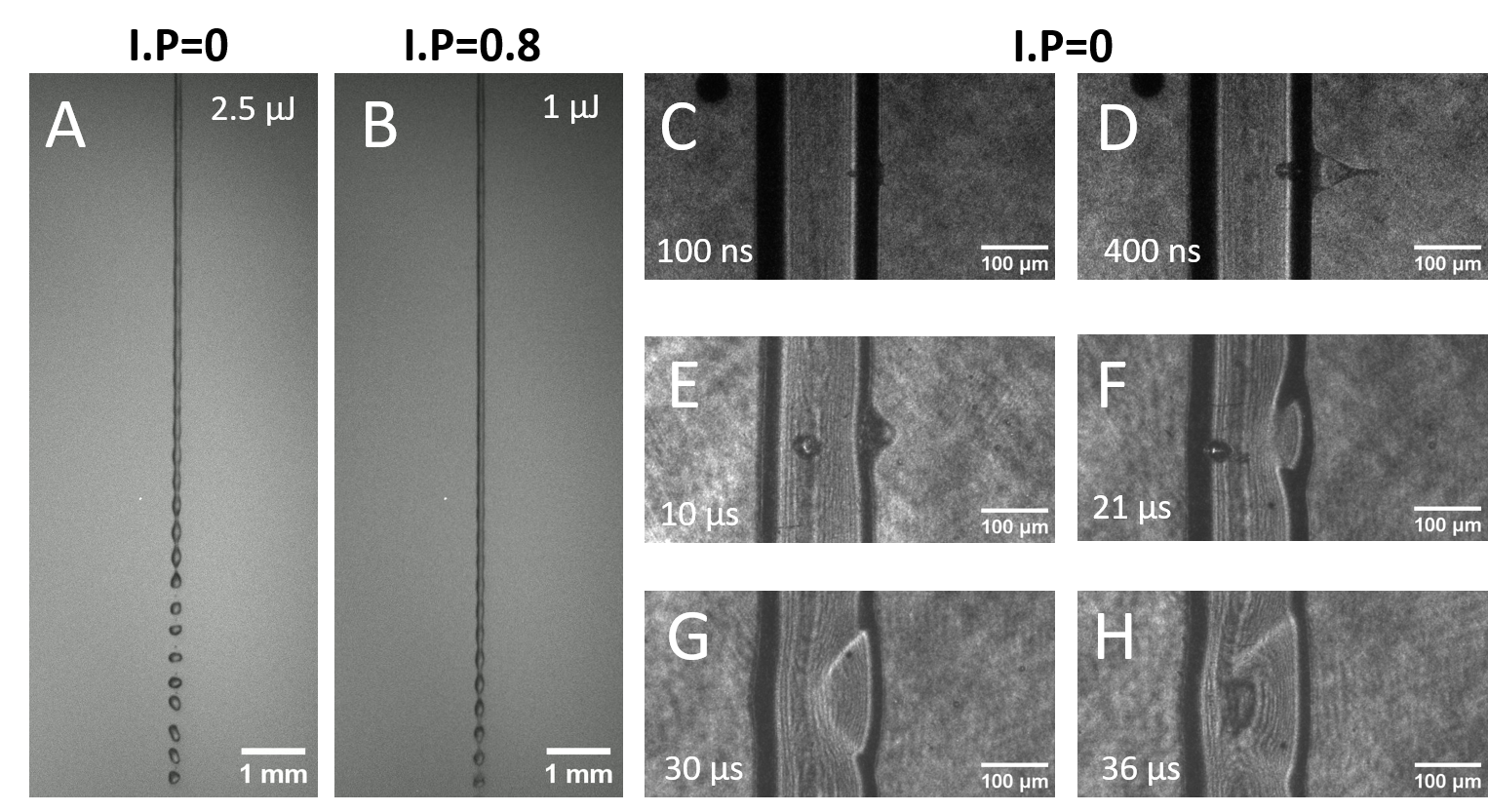} 
\caption{\textbf{Energy coupling via micro-focusing.}
        \textbf{A)} Liquid droplet formation at $2.5~\mu$J for I.P.$=0$, whereas \textbf{B)} shows droplet formation at $1~\mu$J for I.P.$=0.8$.
        \textbf{C)} For I.P.$=0$, micro-focusing at the back surface of the liquid jet creates local heating at the rear side.
        \textbf{D)} Bubble collapse generates a microjet and a recoiling bubble.
        \textbf{E)} The rear surface starts to relax while the front surface becomes perturbed by the recoiling bubble as seen in \textbf{F)}.
        The front surface continues to be perturbed while the rear surface further relaxes in \textbf{G)}, and finally \textbf{H)} shows both front and rear surfaces evolving into non-local perturbations.}
\label{IP_growth_dependence}
\end{figure}

Because the focal spot is smaller than the jet radius, the interaction position can be varied radially. We characterize this by the impact parameter ${\rm I.P.}=d/R$, where $d$ is the distance from the jet center and $R$ is the jet radius. An I.P. of 0.8 produces a higher local intensity at the rear side of the jet compared with I.P.$=0$, and therefore a lower threshold for optical breakdown and droplet generation. As shown in Fig.~\ref{IP_growth_dependence}(A,B), the threshold energy for droplet formation is about $2.5~\mu$J for I.P.$=0$, but only $\sim 1~\mu$J for I.P.$=0.8$. This demonstrates that the efficiency of laser-driven perturbation and breakup depends strongly on the chosen impact parameter. In all experiments reported here, we use I.P.$=0.8$ to minimize the threshold energy and maximize coupling efficiency. This localized rear-interface breakdown initiates the plasma and cavitation dynamics described in the following subsections.

\subsection{ Optical breakdown, plasma formation, and shock generation}

Upon reaching the breakdown threshold, a dense plasma is generated within a volume of characteristic size \(\sim 5\)--\(10~\mu\)m near the rear interface. The plasma formation and initial expansion occur on picosecond to few-nanosecond time scales \cite{Mechanical_Vogel}. In our time-resolved pump--probe images (Fig.~1B1), the plasma appears as a localized region of strong absorption and scattering, reaching its maximum extent by $t_{\rm plasma} \approx 2~\mathrm{ns}$.

The rapid energy deposition drives a rapid expansion that launches a shock wave into the surrounding liquid. The shock is observed as a narrow, high-contrast feature in the shadowgraphy images (Fig.~1B2), with an initial width of $w_{\rm shock} \sim 5~\mu\mathrm{m}$, corresponding to a temporal duration $\tau_{\rm shock} \approx \frac{w_{\rm shock}}{v_s} \approx 3~\mathrm{ns}$, for a shock velocity $v_s \approx 2.0~\mathrm{km/s}$, consistent with gigapascal-level pressures \cite{Gpa_Shock_Nat2016}. The acoustic transit time across the jet is $\tau_{\rm ac} \approx \frac{D}{c_s} \approx \frac{100~\mu\mathrm{m}}{1500~\mathrm{m/s}} \approx 70~\mathrm{ns}$, where $c_s \approx 1500~\mathrm{m/s}$ is the speed of sound in water. Thus, the shock and associated acoustic waves traverse the jet and undergo multiple reflections on time scales $\lesssim 100$~ns.

\subsection{Time-scale separation: shock vs. capillary response}

To assess whether the shock stage can directly seed a lasting capillary perturbation, we compare the relevant time scales to the capillary--inertial response time of the jet. The capillary time is defined as $\tau_{\rm cap} = \sqrt{\frac{\rho R^3}{\sigma}}$, where $\rho \approx 1000~\mathrm{kg/m^3}$ is the liquid density, $\sigma \approx 0.072~\mathrm{N/m}$ is the surface tension, and $R = 50~\mu\mathrm{m}$ is the jet radius. This yields $\tau_{\rm cap} \approx 42~\mu\mathrm{s}$. The Ohnesorge number for the jet is $Oh = \frac{\eta}{\sqrt{\rho \sigma D}} \approx 0.01$, with $\eta \approx 1~\mathrm{mPa\,s}$ the dynamic viscosity, confirming that the surface dynamics are dominated by inertia and surface tension rather than viscosity.

The shock duration (\(\tau_{\rm shock} \sim 3\) ns) and acoustic transit time (\(\tau_{\rm ac} \sim 70\) ns) are more than two orders of magnitude shorter than \(\tau_{\rm cap}\). Consequently, the shock-driven acoustic field undergoes multiple compression--reflection cycles within the jet before the free surface can respond appreciably on the capillary time scale. Consistent with prior work on shock--jet interactions \cite{jet_shock_Nano2020}, we observe no measurable long-lived surface deformation attributable to the shock stage (Fig.~1B3). The shock effectively ``rings'' the jet elastically without imprinting a persistent varicose mode capable of seeding RP growth.

\subsection{ Cavitation bubble dynamics and rear-surface rupture}

The first persistent surface deformation emerges on microsecond time scales through cavitation dynamics. A fraction of the absorbed laser energy is converted into the formation and expansion of a cavitation bubble near the rear interface {\cite{Patrascioiu2014}}. While the exact conversion efficiency depends on the breakdown conditions and local thermodynamics, a conservative estimate based on prior measurements suggests that \(\sim 5\%\) of the absorbed energy is channeled into cavitation \cite{Energy_cavitation}.

For an absorbed energy of $E_{\rm Abs} \sim 100~\mathrm{nJ}$, this yields a cavitation energy $E_{\rm bubble} \sim 5~\mathrm{nJ}$. In an unbounded liquid, this energy would drive approximately spherical bubble expansion to a maximum Rayleigh radius $R_{b,\rm max}$ given by the balance between the bubble energy and the work done against ambient pressure and surface tension. Neglecting ambient pressure for an order-of-magnitude estimate and considering only the surface term, the capillary energy barrier associated with creating a surface deformation of size $R_{b,\rm max}$ is $E_\sigma \approx \pi R_{b,\rm max}^2 \gamma$. For $R_{b,\rm max} \sim 50~\mu\mathrm{m}$ and $\gamma \approx 0.072~\mathrm{N/m}$, this gives $E_\sigma \approx 0.57~\mathrm{nJ}$.

Since $E_{\rm bubble} \gg E_\sigma$, the expanding cavitation bubble carries sufficient energy to easily overcome the capillary barrier associated with deforming the jet surface. In our confined geometry, the bubble therefore expands until it encounters the rear liquid--air interface, typically within a few microseconds after nucleation (Fig.~1B3). The bubble expansion time scale can be estimated from the Rayleigh collapse/expansion time $\tau_{\rm exp} \sim 0.915 R_{b,\rm max} \sqrt{\frac{\rho}{P_\infty}}$, which for $R_{b,\rm max} \sim 50~\mu\mathrm{m}$ and ambient pressure $P_\infty \approx 1$~atm yields $\tau_{\rm exp} \approx 4\text{--}5~\mu\mathrm{s}$, consistent with our observations. Upon reaching the free surface, the bubble ruptures the rear interface (Fig.~1B4), leading to the ejection of a directional microjet into air~\cite{Patrascioiu2014}. This process is analogous to laser-induced forward jetting from free surfaces and thin films, where asymmetric bubble collapse or expansion near an interface produces a focused liquid jet.

\subsection{ Recoil-driven deformation and axisymmetrization}

Rupture and microjet ejection impart a recoil momentum to the remaining liquid column. By conservation of momentum, the unidirectional microjet launch drives an opposite disturbance that propagates across the jet diameter. This recoil pulse traverses the $100~\mu$m jet on a microsecond time scale and deforms the front interface (Fig.~1B5--B7), creating a localized surface indentation.

Although optical breakdown, cavitation growth, rupture, and recoil are strongly asymmetric (localized near the rear interface, with a directed microjet into air), the resulting deformation evolves under capillary--inertial dynamics on the time scale $\tau_{\rm cap}$. Surface tension minimizes the surface area, redistributing momentum around the circumference and damping azimuthal asymmetries. Over $\sim \tau_{\rm cap}$, the initially asymmetric deformation relaxes toward an approximately axisymmetric varicose (necking) mode (Fig.~1B8), the appropriate symmetry for subsequent Rayleigh--Plateau amplification. This axisymmetrization is consistent with the known behavior of localized disturbances on liquid jets, where capillary relaxation filters out higher-order azimuthal modes and leaves the axisymmetric component to dominate the long-time dynamics \cite{Eggers_RMP1997, Eggers_RPP2008}.

\subsection{ From single-pulse response to periodic perturbation train}

When the laser is operated in a periodic, multi-kHz regime, the above sequence repeats at each pulse, generating a train of isolated deformations that are advected downstream with the liquid flow. Because the time between pulses (\(\sim 10\)--\(100~\mu\)s, depending on repetition rate) is comparable to or longer than \(\tau_{\rm cap}\), each perturbation has sufficient time to relax into an axisymmetric RP seed before the next pulse arrives.

The resulting instability therefore inherits two independently controllable parameters from the optical excitation. The spatial separation between successive perturbations is set by the laser repetition rate $f_{\rm rep}$ and the jet velocity $v_{\rm jet}$ via $\lambda = v_{\rm jet}/f_{\rm rep}$, so that $kR = 2\pi R/\lambda$, while the perturbation amplitude is controlled by the deposited pulse energy, which determines the cavitation energy and hence the strength of the recoil-driven deformation. This mechanism provides a reproducible and tunable means of seeding the Rayleigh--Plateau instability with well-defined initial conditions, forming the basis for the systematic exploration of the $(kR,\epsilon/R)$ perturbation landscape presented in the main text.


\section{Stability and size control of wavelength-selected droplet states}
To further characterize the wavelength-selected breakup states discussed in the main text, we quantify the temporal and morphological stability of the resulting droplets and examine how their production and relative sizes can be controlled by the laser repetition rate and the interaction length between successive laser perturbations. These measurements demonstrate reproducible control of liquid-jet breakup from synchronized monodisperse droplet trains to stable bidisperse states, isolated droplets, and controlled droplet-size distributions.

\begin{figure}[h!]
\centering
\includegraphics[width=1.0\textwidth]{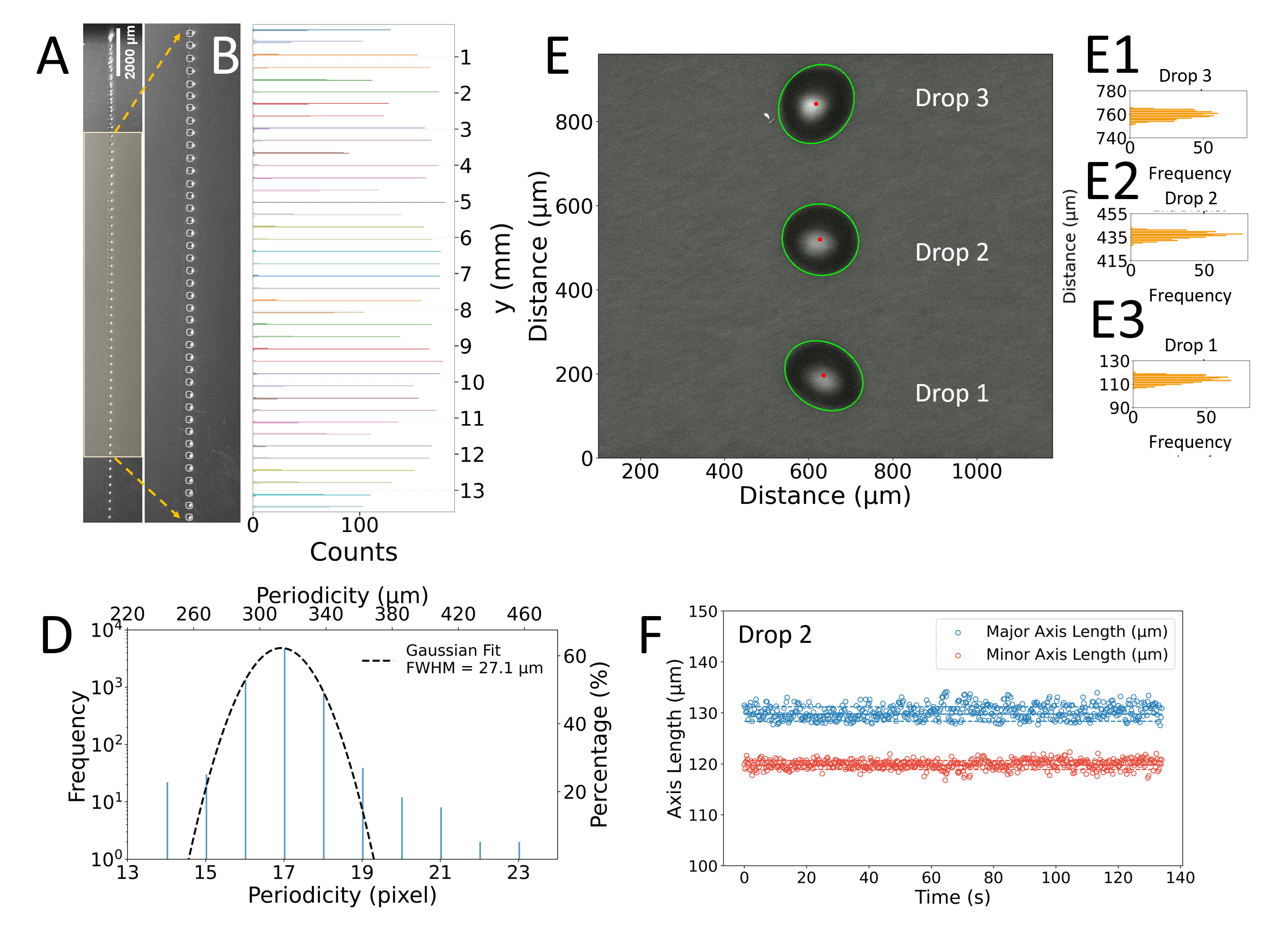}
\caption{\textbf{Temporal and morphological stability of the monodisperse droplet train.}
\textbf{A)} Long-time evolution of the liquid jet showing symmetric breakup into more than 50 droplets.
\textbf{B)} Representative sequence of 40 droplets selected for quantitative analysis.
\textbf{C)} Center positions of the droplets tracked over time to quantify temporal fluctuations and spatial jitter.
\textbf{D)} Distribution of the spacing between adjacent droplets, characterizing the spatial regularity of the train.
\textbf{E)} Magnified view of a representative region of the train.
\textbf{F)} Temporal evolution of a representative droplet used to quantify morphological fluctuations.}
\label{dropletstability}
\end{figure}

We first consider the monodisperse state realized near the fastest-growing Rayleigh--Plateau mode, $kR\simeq0.9$ ($\lambda/D\simeq3.5$), where periodic laser perturbations produce a synchronized droplet train (main text, Fig.~5A). A sequence of 40 successive droplets is analyzed over approximately $47~\mathrm{s}$ (Fig.~\ref{dropletstability}A--C). The centroid fluctuations have a standard deviation of approximately $11~\mu\mathrm{m}$, comparable to one pixel of the imaging system, while the inter-droplet spacing remains highly regular (Fig.~\ref{dropletstability}D). A magnified view of the train is shown in Fig.~\ref{dropletstability}E. The morphological stability is quantified from the temporal evolution of a representative droplet (Fig.~\ref{dropletstability}F), whose major and minor axes are $183~\mu\mathrm{m}$ ($\sigma=2~\mu\mathrm{m}$) and $169~\mu\mathrm{m}$ ($\sigma=1~\mu\mathrm{m}$), respectively. Thus, the synchronized monodisperse state exhibits both long-time positional and morphological stability.

\begin{figure}[h!]
\centering
\includegraphics[width=1.0\textwidth]{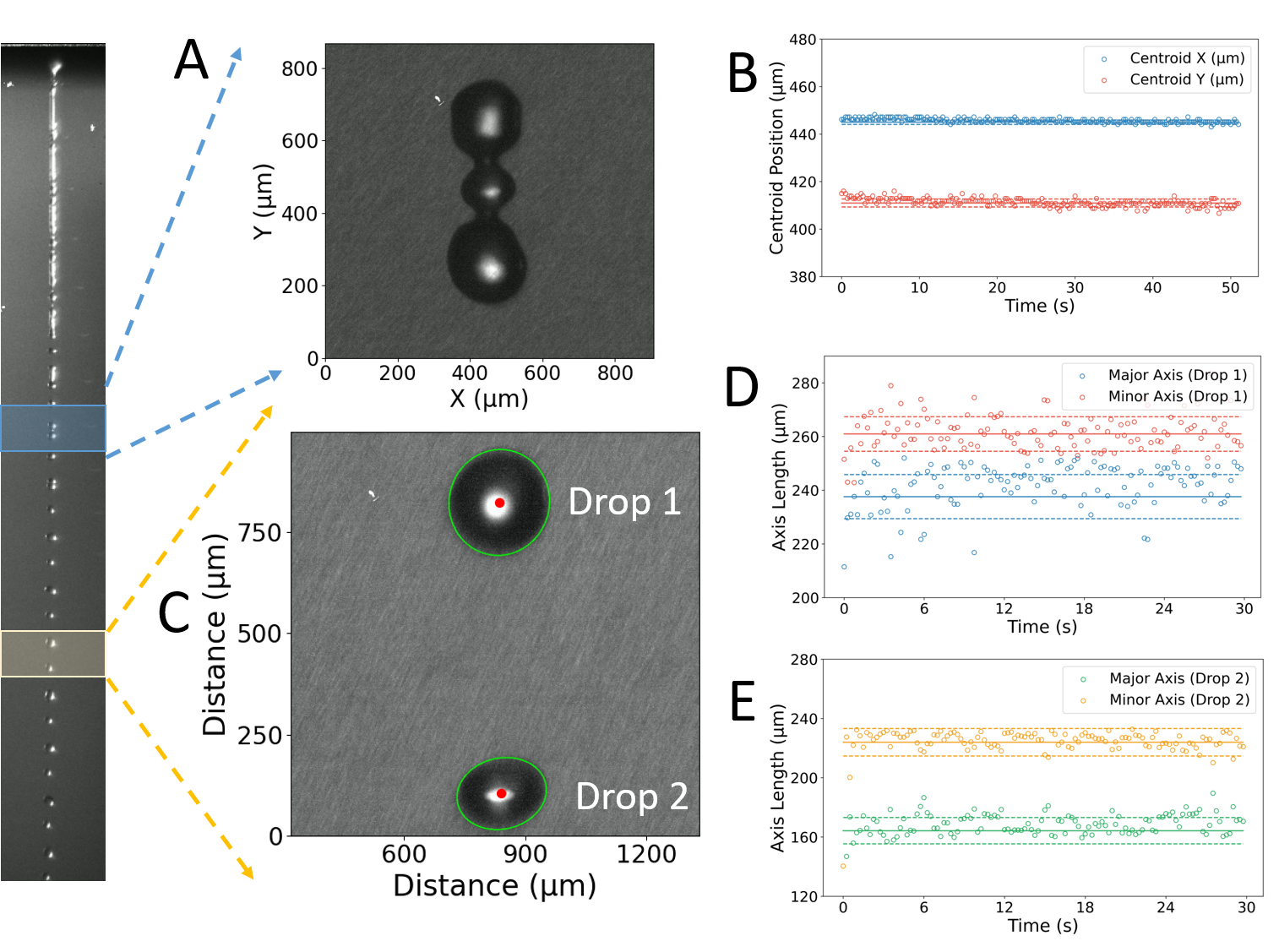}
\caption{\textbf{Stability and size contrast in the bidisperse breakup regime.}
\textbf{A)} Localized deformation and dumbbell-like intermediate structure at $kR=0.26$ ($\lambda/D\simeq10$).
\textbf{B)} Temporal evolution of the intermediate structure prior to breakup.
\textbf{C)} Final bidisperse configuration, with Drop~2 originating from the laser-perturbed region and Drop~1 from the intervening unperturbed segment.
\textbf{D,E)} Temporal evolution of the major and minor axes of the two droplet populations.}
\label{localizedperturbation}
\end{figure}

At the longer perturbation wavelength, $kR=0.26$ ($\lambda/D\simeq10$), the breakup becomes bidisperse (main text, Fig.~5B). The localized cavitation-driven deformation develops into the smaller laser-seeded droplet (Drop~2), while the intervening unperturbed liquid segment retracts under surface tension to form the larger droplet (Drop~1) (Fig.~\ref{localizedperturbation}A--C). The intermediate dumbbell-like structure persists over several seconds before final fragmentation. Over an observation time exceeding $30~\mathrm{s}$, both droplet populations remain morphologically stable. Drop~2 has major and minor axes of $224~\mu\mathrm{m}$ ($\sigma\approx9~\mu\mathrm{m}$) and $164~\mu\mathrm{m}$ ($\sigma\approx9~\mu\mathrm{m}$), respectively, whereas Drop~1 has corresponding axes of $261~\mu\mathrm{m}$ ($\sigma\approx6~\mu\mathrm{m}$) and $238~\mu\mathrm{m}$ ($\sigma\approx8~\mu\mathrm{m}$) (Fig.~\ref{localizedperturbation}D,E). Thus, increasing the perturbation wavelength provides deterministic control over the breakup morphology, switching from a highly regular monodisperse train to a stable bidisperse state with distinct droplet sizes.

The localized nature of the laser-induced perturbation also allows the breakup process to be controlled over a broad range of repetition rates, from isolated perturbations to synchronized sub-MHz droplet production.

\begin{figure}[h!] 
\centering
\includegraphics[width=0.6\textwidth]{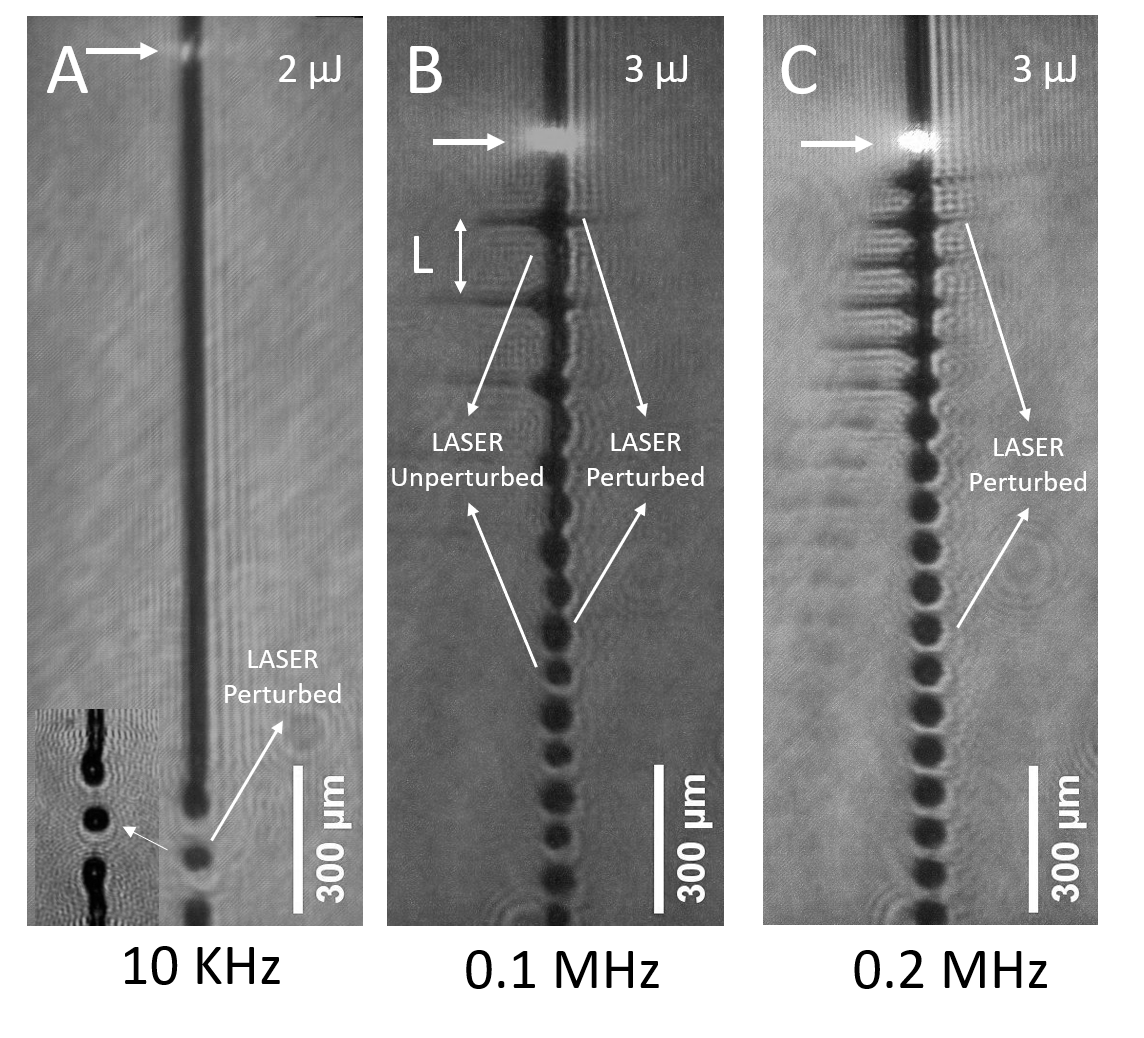}
\caption{\textbf{Localized perturbation and synchronized droplet generation at increasing laser repetition rates.}
\textbf{A)} Local perturbation of the liquid jet at a $10~\mathrm{kHz}$ laser repetition rate, where a droplet forms at the laser interaction point after approximately $100~\mu\mathrm{s}$.
\textbf{B)} Bidisperse droplet generation at a $0.1~\mathrm{MHz}$ repetition rate.
\textbf{C)} Droplet formation at a sampling rate of $0.2~\mathrm{MHz}$, matching the laser repetition rate.}
\label{local_to_multimhz}
\end{figure}

At a $10~\mathrm{kHz}$ repetition rate, a localized laser interaction produces an individual droplet at the laser interaction point after approximately $100~\mu\mathrm{s}$ (Fig.~\ref{local_to_multimhz}A). As the repetition rate is increased to $0.1~\mathrm{MHz}$, the shorter interval between successive perturbations leaves an intervening segment of liquid that undergoes secondary capillary breakup, producing a bidisperse droplet population (Fig.~\ref{local_to_multimhz}B). At $0.2~\mathrm{MHz}$, droplet formation becomes synchronized with the laser repetition rate (Fig.~\ref{local_to_multimhz}C), demonstrating controlled droplet production at rates up to $0.2~\mathrm{MHz}$. These observations further show that the same localized perturbation mechanism can be operated from isolated single-droplet formation to high-rate synchronized droplet generation.

The relative sizes of the droplets can be further controlled through the volume of liquid transported between successive laser interactions. For a cylindrical jet of radius $r$, velocity $v$, and laser repetition rate $f$, the volume transported between successive pulses is $V_{\rm seg}=\pi r^2 v/f$. If the localized interaction extracts an approximately constant volume $V_c$ to form the laser-seeded droplet (Drop~2), the remaining volume forming Drop~1 is $V_1=\pi r^2 v/f - V_c$. Thus, for fixed interaction conditions, $V_2\simeq V_c$ remains approximately constant, whereas the volume of Drop~1 increases with the pulse interval $T=1/f$ or, equivalently, with the axial interaction length $L=v/f$.

\begin{figure}[h!]
\centering
\includegraphics[width=1\textwidth]{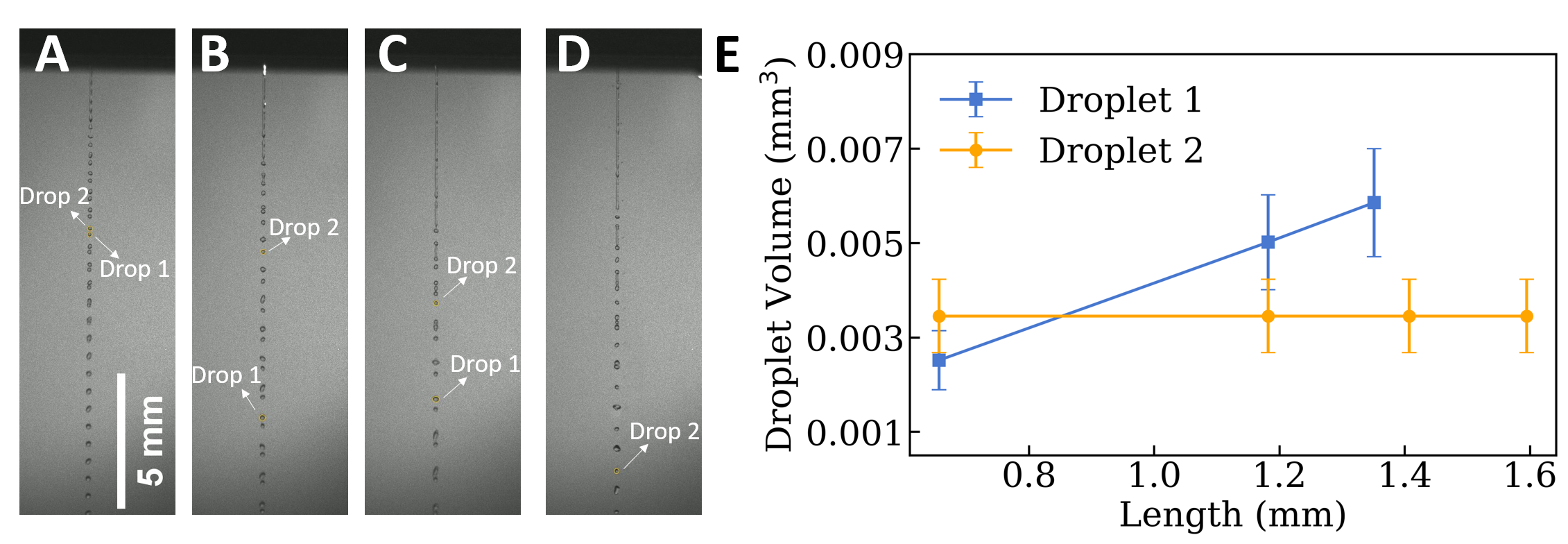}
\caption{\textbf{Volume scaling and droplet-size control through the interaction length.}
\textbf{A--D)} Liquid-jet breakup following successive laser-pulse interactions at different jet velocities. Varying the jet velocity changes the axial interaction length between successive pulses, thereby controlling the volume of the resulting droplets.
\textbf{E)} Variation of the droplet volume with interaction length for two successive droplets. Drop~1 increases in volume with interaction length, whereas Drop~2 remains approximately constant, indicating that the interaction length provides a means to control the resulting droplet size.}
\label{volumescaling_rep}
\end{figure}

The predicted volume scaling is confirmed experimentally by varying the jet velocity while maintaining the laser repetition rate (Fig.~\ref{volumescaling_rep}A--D). Increasing the jet velocity increases the axial distance between successive laser interactions and therefore the volume of liquid transported between them. The volume of Drop~1 increases systematically with interaction length, whereas the volume of Drop~2 remains approximately constant (Fig.~\ref{volumescaling_rep}E). This behavior is consistent with the volume-conservation picture above and demonstrates an additional degree of control over the relative sizes of the two droplet populations.

Taken together, these measurements demonstrate that localized femtosecond-laser seeding provides reproducible control over liquid-jet breakup across multiple scales. The perturbation wavelength controls the breakup morphology, producing stable monodisperse or bidisperse states; the laser repetition rate controls the spatial and temporal sequence of perturbations, enabling operation from isolated droplets to synchronized production at $0.2~\mathrm{MHz}$; and the interaction length provides quantitative control over the relative droplet volumes. Thus, the same localized seeding mechanism provides control not only over whether and when the jet breaks up, but also over the resulting droplet morphology, spacing, and size.

For completeness, we note that within linear stability theory a localized surface perturbation decomposes into a continuum of Fourier modes that each evolve according to the classical Rayleigh--Plateau dispersion relation; the shape of the initial bump affects only the modal amplitudes, not the dispersion relation itself~\cite{Chandrasekhar1961,Eggers_RMP1997}. Solving the linearized kinematic and dynamic (Bernoulli) boundary conditions mode-by-mode gives
\begin{equation}
\omega_n^2 \;=\; \frac{\sigma}{\rho}\,k_n\Big(\frac{1}{R_0^2}-k_n^2\Big)
\frac{I_1(k_n R_0)}{I_0(k_n R_0)},
\label{eq:dispersion-box}
\end{equation}
where $I_0$ and $I_1$ are modified Bessel functions of the first kind. Maximizing Eq.~(\ref{eq:dispersion-box}) yields $k_{\rm max}R_0 \approx 0.697$ ($\lambda_{\rm max} \sim 9\,R_0$), and equating the volume of this cylindrical segment to that of the resulting droplet gives $R_d \sim 1.88\,R_0$. This estimate provides a rough consistency check for the droplet sizes in the weakly perturbed, near-linear regime, but the primary focus of this work is the controlled access to nonlinear and beyond-stability breakup via finite-amplitude laser seeding.


\section{Strong-seed perturbation: amplitude control and breakup response}

This section collects measurements that characterize how the strength of the laser-induced seed perturbation controls the subsequent breakup dynamics. We quantify (i) how the surface deformation amplitude grows with deposited pulse energy, (ii) how the breakup length shifts with energy at fixed repetition rate, and (iii) how breakup length and axial mode spectra evolve with perturbation frequency at high pulse energy. Together, these data establish the laser pulse energy and repetition rate as independent control knobs for the initial perturbation amplitude and wavenumber, and they provide the experimental basis for the strongly perturbed and beyond-Rayleigh--Plateau regimes discussed in the main text.

\begin{figure}[h!]
\centering
\includegraphics[width=0.9\textwidth]{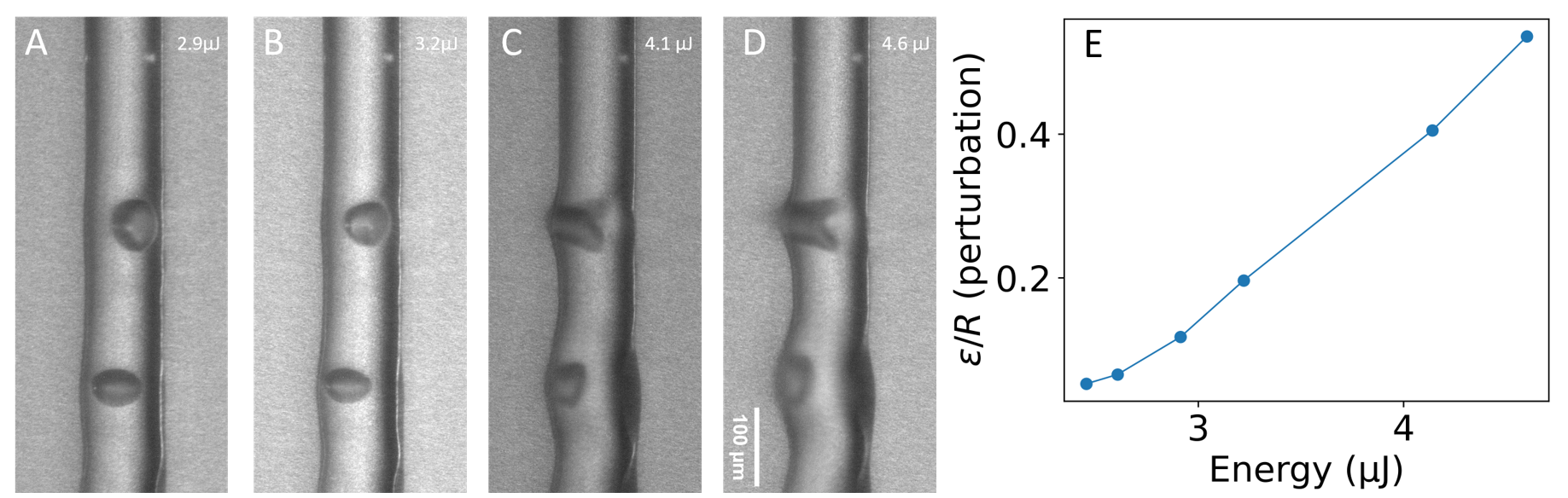} 
\caption{\textbf{Perturbation amplitude control via laser pulse energy.}
(\textbf{A}--\textbf{D}) Increase of the surface perturbation amplitude with increasing laser pulse energy.
(\textbf{E}) Quantitative evolution of the liquid-jet surface perturbation, evaluated $40~\mu$s after energy deposition, as a function of the deposited energy.}
\label{energy_perturbation_strength}
\end{figure}

Figure~\ref{energy_perturbation_strength} quantifies the relation between deposited pulse energy and the resulting surface perturbation strength. Panels (A)--(D) show the growth of the interfacial deformation with increasing energy, while panel (E) plots the perturbation amplitude measured at a fixed delay of $40~\mu$s after energy deposition. This monotonic increase establishes pulse energy as the primary control parameter for the initial perturbation amplitude, and hence for the transition from weak, linear Rayleigh--Plateau-like growth to strongly nonlinear, laser-dominated breakup.

\begin{figure}[h!]
\centering
\includegraphics[width=1\textwidth]{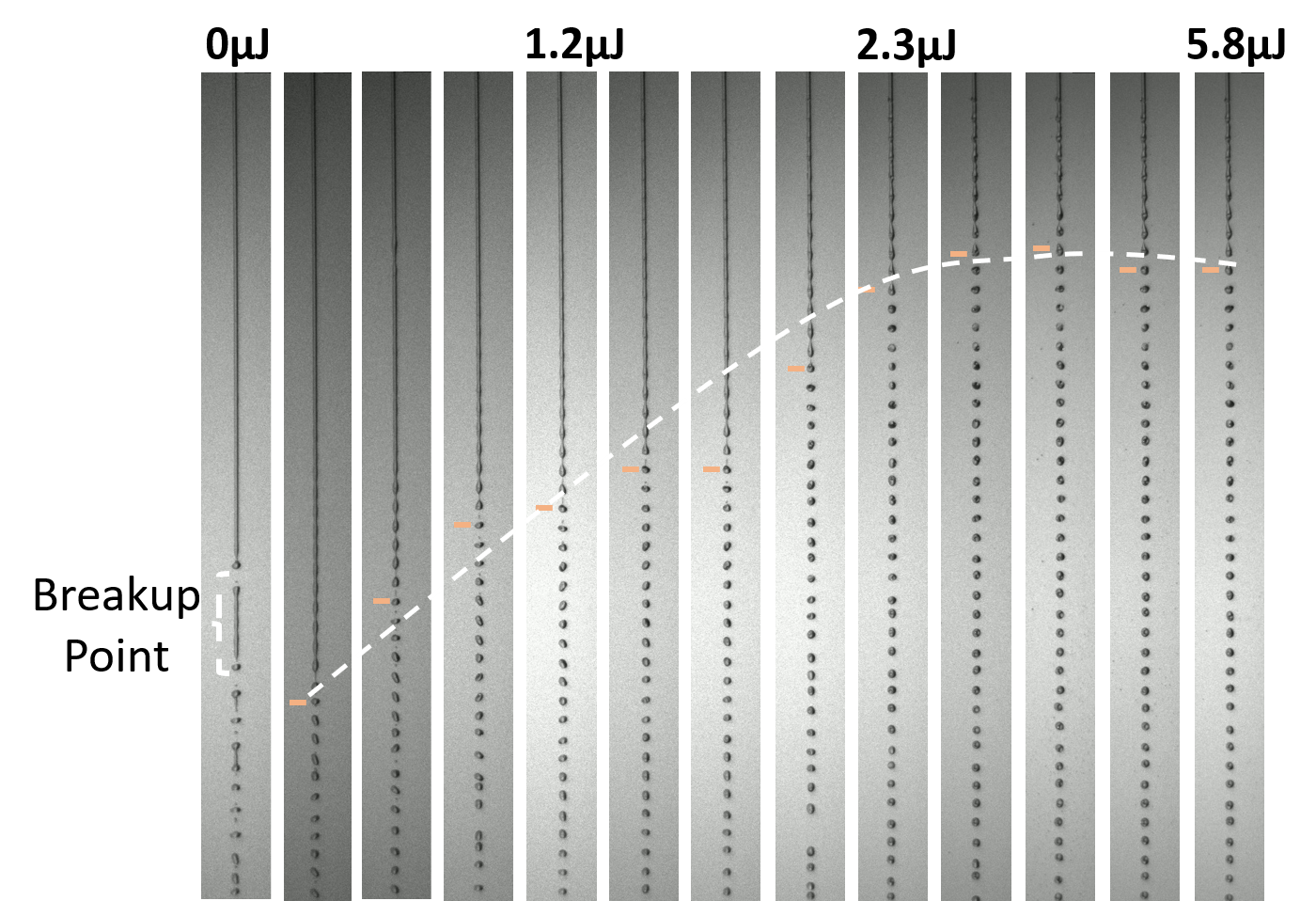} 
\caption{\textbf{Energy-dependent breakup length at the resonance frequency.}
Pictorial representation of the change in breakup length of the liquid jet at the resonance frequency (25~kHz) as a function of laser pulse energy. Starting from the natural breakup observed at 0~\textmu J, the position at which the first droplet appears shifts systematically with increasing energy.}
\label{breakup_reversal}
\end{figure}

At fixed repetition rate, increasing the perturbation amplitude shortens the distance required for the jet to break up. Figure~\ref{breakup_reversal} shows the breakup length at the resonance frequency (25~kHz) as a function of pulse energy, starting from the natural (unforced) breakup at 0~\textmu J. With increasing energy, the first droplet appears progressively closer to the interaction point, reflecting stronger initial deformations and faster onset of pinch-off. This behavior is consistent with the amplitude-controlled transition from linear to nonlinear breakup observed in the main text.

\begin{figure}[h!]
\centering
\includegraphics[width=0.8\textwidth]{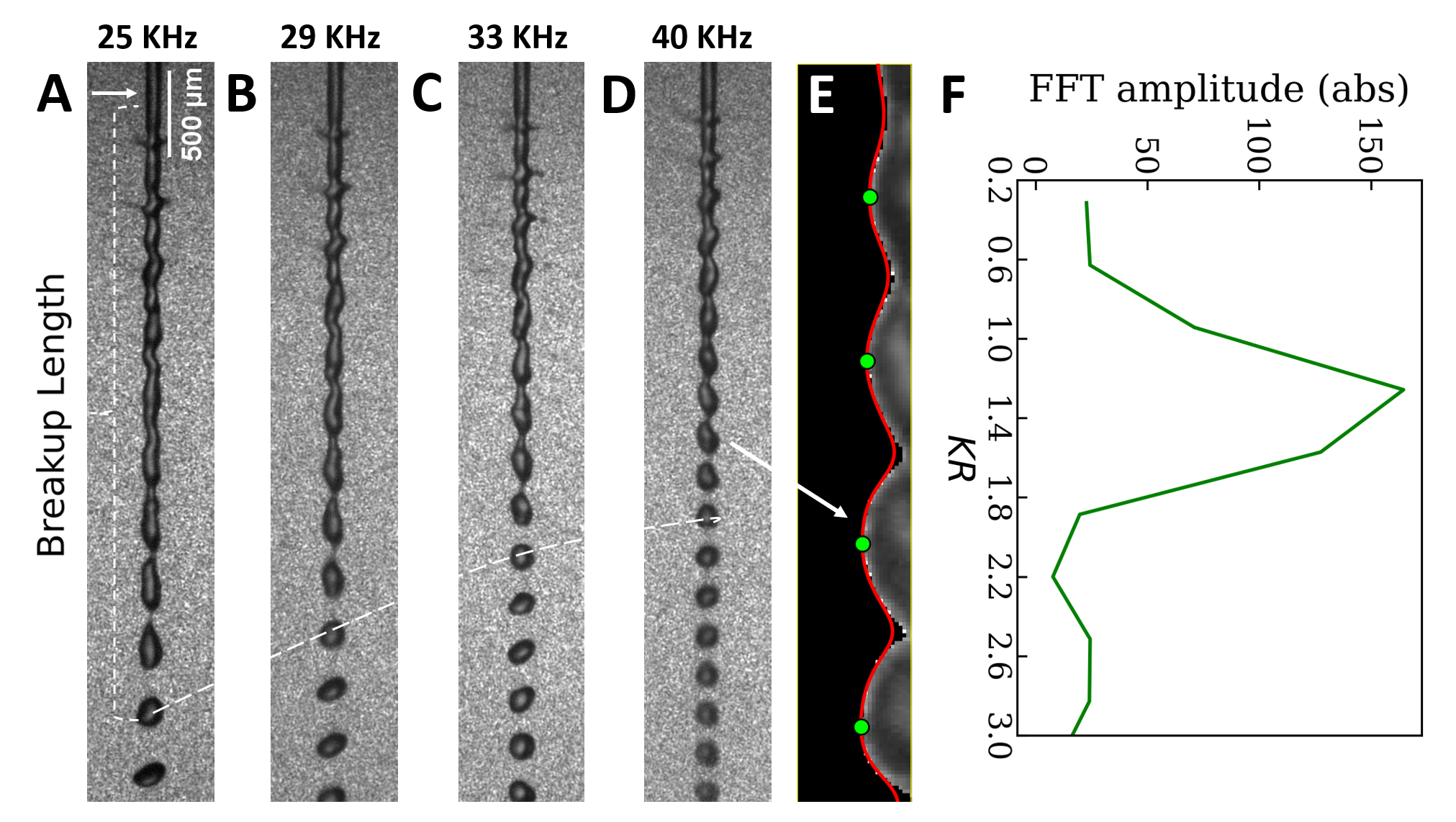} 
\caption{\textbf{Breakup length and mode spectrum at high perturbation strength.}
(\textbf{A}--\textbf{D}) Breakup length as a function of perturbation frequency at a laser pulse energy of $2.4~\mu$J.
(\textbf{F}) Outline of the left-side liquid jet recorded at a repetition rate of 40~kHz.
(\textbf{G}) Fast Fourier transform (FFT) of the jet outline shown in (\textbf{F}), revealing the excited axial modes.}
\label{jet_fmscan_higher_energy}
\end{figure}

To probe the response at high perturbation strength across different wavenumbers, we fix the pulse energy at $2.4~\mu$J and scan the repetition rate. Figure~\ref{jet_fmscan_higher_energy}(A)--(D) shows the resulting variation of breakup length with perturbation frequency. At these energies, the jet response is strongly nonlinear, and the breakup length no longer follows the simple linear Rayleigh--Plateau expectation. Panels (F) and (G) illustrate the analysis procedure: the jet outline is extracted (F) and its axial structure is decomposed via FFT (G) to identify the dominant excited modes. These spectra confirm that, at high energy, modes with $kR>1$ are robustly excited, providing the experimental basis for the beyond-stability breakup reported in the main text.

For the dispersion relation shown in Fig.~2B of the main text, the dimensionless wavenumber is obtained from the repetition rate and jet velocity as $kR = 2\pi R f_{\rm rep}/v_{\rm jet}$, and the temporal growth rate $\omega$ is extracted by fitting the envelope of the perturbation amplitude $a(z)$ in the linear growth regime to $a(z)=a_0\exp(\omega z/v_{\rm jet})$. The normalized growth rate $\omega^*=\omega\sqrt{\rho R_0^3/\sigma}$ is then used to construct the dispersion relation. The classical inviscid Rayleigh--Plateau relation is $\omega^2 = (\sigma/\rho)\,k(1/R_0^2-k^2)\,I_1(kR_0)/I_0(kR_0)$.


\section{Breakup beyond the Rayleigh--Plateau stability limit}

The main text establishes that, above a threshold energy $E_c\approx2.18~\mu$J, the jet response departs from linear Rayleigh--Plateau growth and that strong local perturbations can drive breakup for $kR>1$, where infinitesimal perturbations are linearly stable (main text, Fig.~4). This section provides the detailed growth- and decay-rate measurements, mode spectra, and additional wavenumber scans that support those claims but are too extensive for the main text. In particular, we quantify (i) the transition from Rayleigh--Plateau-like growth to nonlinear, laser-driven breakup, (ii) the decay of linearly stable modes at moderate energy, and (iii) the recovery of deterministic breakup at still higher energies and $kR\gg1$.

\begin{figure}[h!]
    \centering
    \includegraphics[width=\linewidth]{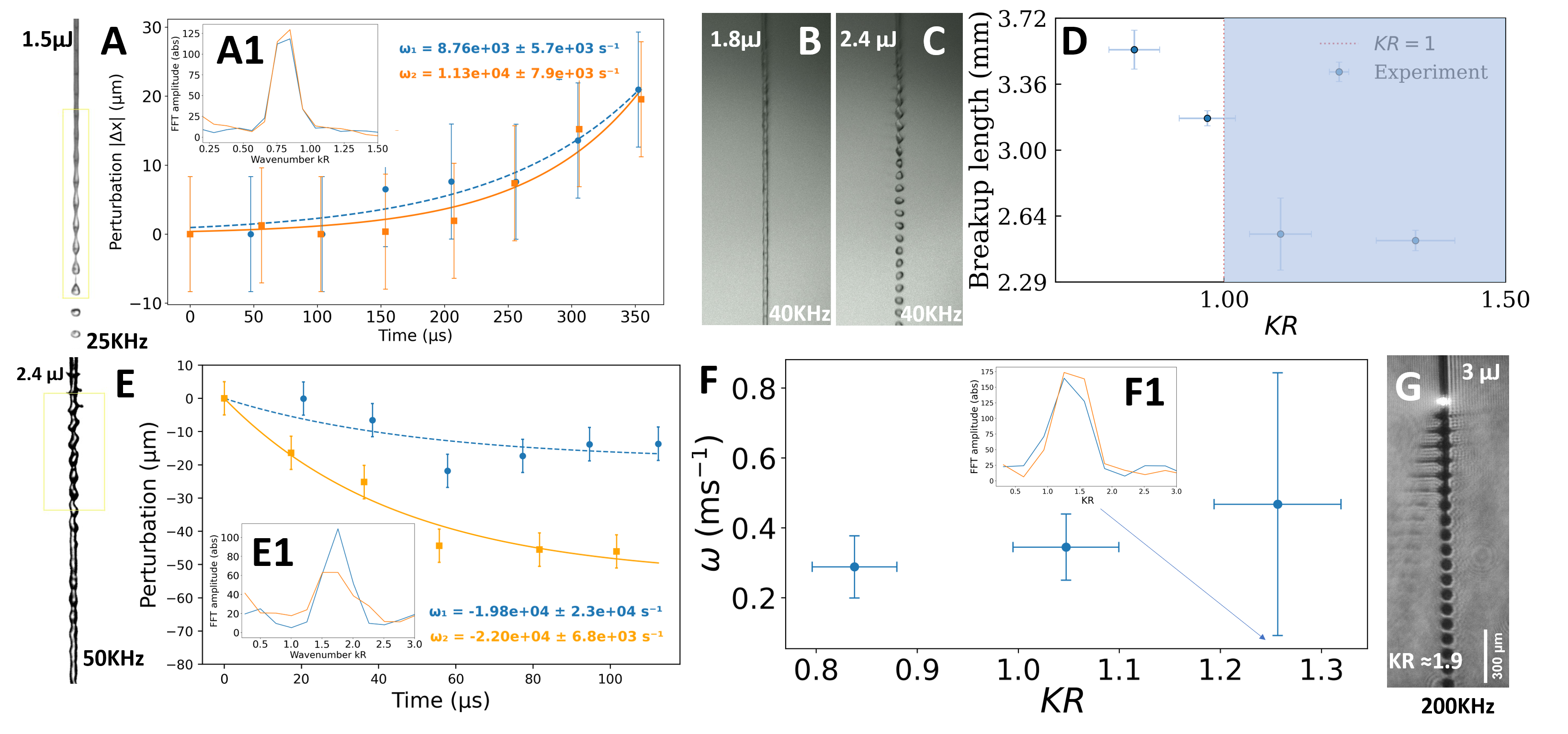}
    \caption{
    \textbf{Beyond Rayleigh--Plateau instability.} 
    (\textbf{A}) Interaction of a liquid jet with a 1.5~\textmu J laser pulse, showing jet breakup dominated by the mode with dimensionless wavenumber $kR = 0.8$ (A1, inset). The temporal growth of the surface perturbation is quantified by fitting an exponential to the measured amplitude evolution on both sides of the jet, yielding growth rates of $8.76\times10^{3}\,\mathrm{s^{-1}}$ and $1.13\times10^{4}\,\mathrm{s^{-1}}$, respectively.
    (\textbf{B}) Liquid jet interaction with a laser pulse energy of 1.8~\textmu J at 40~kHz, showing no jet breakup. 
    (\textbf{C}) Under otherwise identical conditions but with an increased pulse energy of 2.4~\textmu J, the jet undergoes breakup, forming droplets. 
    (\textbf{D}) At an energy of $2.4~\mu$J (high perturbation strength), the interaction demonstrates access to regimes beyond the Rayleigh--Plateau limit.
    (\textbf{E}) Interaction of a liquid jet with a 2.4~\textmu J laser pulse, showing surface perturbations dominated by modes with wavenumber $kR = 1.7$ and $kR = 1.5$ (E1, inset). The temporal decay of the surface perturbation is quantified by fitting an exponential to the measured amplitude evolution on both sides of the jet, yielding decay rates of $1.98\times10^{4}\,\mathrm{s^{-1}}$ and $2.2\times10^{4}\,\mathrm{s^{-1}}$, respectively.
    (\textbf{F}) Absolute growth rate of the surface perturbation as a function of the dimensionless wavenumber $kR$, determined from the Fourier spectrum (F1).
    (\textbf{G}) Droplet formation at 0.2~MHz repetition rate, corresponding to wavenumber $kR \approx 1.9$, achieved by increasing the perturbation energy to 3~\textmu J.} 
    \label{stronglyperturbed}
\end{figure}

At a pulse energy of $1.5~\mu$J, the jet breakup is dominated by a mode with $kR\approx0.8$, consistent with the linearly unstable Rayleigh--Plateau band (Fig.~\ref{stronglyperturbed}A,A1). Fitting an exponential to the temporal evolution of the surface perturbation on both sides of the jet yields growth rates of $8.76\times10^{3}~\mathrm{s^{-1}}$ and $1.13\times10^{4}~\mathrm{s^{-1}}$, respectively.

Increasing the perturbation amplitude by raising the pulse energy provides access to breakup dynamics that are no longer constrained by the linear Rayleigh--Plateau stability boundary. At a repetition rate of $40~\mathrm{kHz}$ and a pulse energy of $1.8~\mu$J, the jet remains stable and no droplets are formed (Fig.~\ref{stronglyperturbed}B). Under otherwise identical conditions, increasing the energy to $2.4~\mu$J induces breakup and droplet formation (Fig.~\ref{stronglyperturbed}C), demonstrating that sufficiently strong localized forcing can drive breakup even when the linear theory predicts stability.

Figure~\ref{stronglyperturbed}D summarizes the breakup length as a function of the driver wavenumber $kR$ in this high-energy regime, showing droplet formation for $kR>1$ (see also Fig.~\ref{jet_fmscan_higher_energy}). At the same pulse energy of $2.4~\mu$J but a higher repetition rate of $50~\mathrm{kHz}$, corresponding to $kR=1.5$--$1.7$, the jet does not break up; instead, the surface perturbation decays with an effective rate of $\sim2\times10^{4}~\mathrm{s^{-1}}$ (Fig.~\ref{stronglyperturbed}E,E1). Fourier analysis confirms the presence of modes with $kR=1.5$ and $kR=1.7$ on both sides of the jet. The resulting growth (or decay) rates extracted from the $kR$-dependent amplitude measurements (Fig.~\ref{stronglyperturbed}F,F1) are approximately two orders of magnitude smaller than those in the droplet-forming regime, consistent with linearly stable modes that are weakly excited.

Nevertheless, by further increasing both the repetition rate and the pulse energy, droplet generation can be recovered deep in the $kR>1$ regime. At a repetition rate of $0.2~\mathrm{MHz}$ ($kR\approx1.9$) and a pulse energy of $3~\mu$J, deterministic breakup and droplet formation are observed (Fig.~\ref{stronglyperturbed}G). Thus, varying the laser pulse energy controls the initial perturbation amplitude and allows us to probe breakup well beyond the classical Rayleigh--Plateau stability curve, accessing a strongly perturbed, nonlinear regime in which droplet formation is governed by finite-amplitude dynamics rather than linear instability.

\end{document}